\documentclass[aps,prb,twocolumn,superscriptaddress]{revtex4-2}

\usepackage{xcolor} 
\usepackage{graphicx}
\usepackage{dcolumn}
\usepackage{bm}
\usepackage{amsmath}
\usepackage{amssymb} 
\usepackage{braket}
\usepackage{array}
\usepackage{comment}

\usepackage[font=small, justification=centerlast]{caption}
\usepackage{hyperref}

\newcommand\numberthis{\addtocounter{equation}{1}\tag{\theequation}}

\begin{document}


\title{Quench dynamics in higher-dimensional Holstein models: Insights from Truncated Wigner Approaches }


\author{Eva Paprotzki}%
\email{eva.paprotzki@uni-hamburg.de}
\affiliation{I.~Institute for Theoretical Physics, Universität Hamburg, Notkestraße 9-11 - 22607 Hamburg, Germany}%
\author{Alexander Osterkorn}%
\email{alexander.osterkorn@ijs.si}
\affiliation{%
 Jožef Stefan Institute, Jamova 39 - 1000 Ljubljana, Slovenia
}
\author{Vibhu Mishra}
\affiliation{Institute for Theoretical Physics, Georg-August-Universität Göttingen,\\
Friedrich-Hund-Platz 1 - 37077 Göttingen, Germany
}%
\author{Stefan Kehrein}
\affiliation{Institute for Theoretical Physics, Georg-August-Universität Göttingen,\\
Friedrich-Hund-Platz 1 - 37077 Göttingen, Germany
}%

\date{\today}


\begin{abstract}
Charge-density wave phases in quantum materials stem from the complex interplay of electronic and lattice degrees of freedom. Nowadays, various time-resolved spectroscopy techniques allow to actively manipulate such phases and monitor their dynamics in real time.
Modeling such nonequilibrium dynamics theoretically is a great challenge and exact methods can usually only treat a small number of atoms and finitely many phonons.
We approach the melting of charge-density waves in a Holstein model after a sudden switch-on of the electronic hopping from two perspectives: We prove that in the non-interacting and in the strong-coupling limit, the CDW order parameter on high-dimensional hypercubic lattices obeys a factorization relation for long times, such that its dynamics can be reduced to the one-dimensional case.
Secondly, we present numerical results from semiclassical techniques based on the Truncated Wigner Approximation for two spatial dimensions. A comparison with exact data obtained for a Holstein chain shows that a semiclassical treatment of both the electrons and phonons is required in order to correctly describe the phononic dynamics. This is confirmed, in addition, for a quench in the electron-phonon coupling strength.

\end{abstract}

\maketitle

\section{Introduction}

Electron-phonon coupled systems exhibit a variety of different and intriguing phenomena. Though the mechanism of charge-lattice coupling is well known, the consequences of this fundamental interaction are still a topic in present research. 
On one hand, electron-phonon coupling
can lead to Cooper instability, which causes conventional superconductivity,
while it can also cause polaronic charge density waves (CDWs). 
These different ordered phases may compete, or co-exist, like the simultaneous occurence of a superconducting and CDW state in $1T$-$\rm TaS_2$~\cite{Sipos2008}. 
This variety of phenomena makes electron-phonon coupled systems an interesting playground, both theoretically and experimentally, with the ultimate goal to control material properties, e.g.~through optical driving. 
For example, progress in time-resolved spectroscopies enables the imaging of the CDW order parameter through an optical-induced structural phase transition on femto-second time scales \cite{Schmitt2008}\cite{Danz2021}\cite{Eichberger2010}. 
Another such optically-driven transition can be observed in certain copper oxides \cite{Fausti2011} and alkali-doped fullerides \cite{Budden2021}, where a transient metastable state with strongly enhanced conductivity can be generated. 
The role of electron-phonon coupling, especially by means of non-linear processes, in these light-induced, and also in high temperature superconducters is still under debate \cite{Babadi2017}\cite{Murakami2017}\cite{Kemper2017}. 
With experimental methods, which are able to track these ultra-fast nonequilibrium dynamics, a theoretical perspective on that in realistic materials, beyond one-dimensional models, is needed.

On the theoretical side, photo-induced melting of CDW phases has been calculated with exact quantum methods only in one spatial dimension \cite{Hashimoto2017} or small system sizes \cite{DeFilippis2012}. 
Two- or three-dimensional nonequilibrium studies of electron-phonon systems focus on the dilute limit of one or two polarons \cite{Kalosakas1998}\cite{Voulgarakis2000} or have to rely on approximations, like a classical treatment of the phonons \cite{Miyashita2010}\cite{Yonemitsu2009}\cite{Weber2022}. In general, the
separation of energy, and hence time scales, between the
electrons and phonons, motivates the semiclassical approximation. Recent methodological progress 
has been made by adding stochasticity to the classical equations of motion, either by sampling of the initial phononic state from a probability distribution \cite{tenBrink2022}\cite{Weber2022} or by including a self-consistent stochastic noise term in a nonequilibrium dynamical mean field (DMFT) setup \cite{Picano2023}. Our approach belongs to the former type:
We employ the Truncated Wigner Approximation (TWA), which defines an ensemble of mean-field trajectories that retain more quantum features of the initial state than the mean-field approximation.
We implement TWA for the phonons alone as well as TWA for both the phonons and the electrons.
Both approximations go beyond a purely classical description of the lattice.
The first method is also known under the name “multi-trajectory Ehrenfest approach” (see \cite{tenBrink2022} for a comprehensive review).
The second method has -- to our knowledge -- not been applied before.

Semiclassical simulation methods are less restricted by the system size and therefore allow -- within their range of validity -- to study dynamics in higher-dimensional models.
In general, dimensionality can strongly influence physical properties of correlated electron systems, e.g.~the appearance of Luttinger liquids in one spatial dimension or the reduction to an impurity model in infinite dimensions as described by DMFT.
In this paper we ask the question how dimensionality influences the melting of a change-density wave state.
As opposed to optically-driven melting\cite{DeFilippis2012}\cite{Yonemitsu2009}, we consider a quench scenario, where the initial state had been prepared as the ground state to another set of model parameters by a previous protocol.
We start from the observation that the CDW order parameter exactly factorizes at all times if electrons and phonons are uncoupled.
We provide analytical estimates showing that this relation is also satisfied in the strong-coupling regime up to exponentially long times.
We confirm these results numerically in the weak- and strong-coupling regime using the TWA approaches. Thus, this factorization relation can offer a reinterpretation of previous results on CDW melting in one-dimensional systems, in terms of their implications for higher-dimensional geometries.
A direct comparison to exact time evolution data obtained with matrix product state techniques~\cite{Stolpp2020} indicates that -- in particular in the weak-coupling regime -- the additional semiclassical treatment of the electrons improves the agreement with the exact reference data.
In order to shed more light on this aspect, we consider a quench in a 3D Holstein model from zero electron-phonon coupling to weak coupling.
For this problem it becomes particularly clear that phonon-only TWA neglects electronic correlations of the initial state, which immediately cause unphysical behavior.

This paper is structured as follows:\\
The Holstein model is introduced in the following, Section \ref{sec:Holstein-model}, with special emphasis on the initial CDW state because the accuracy of the semiclassical approximations depends on it strongly. Afterwards, we explain the implementation of the semiclassical dynamics via an equation-of-motion description for the TWA and fTWA. Sections \ref{sec:CDW-melting} and \ref{sec:interaction_quench} discuss the results for the CDW melting with TWA and fTWA+TWA and the interaction quench, respectively. Reasons for shortcomings of the TWA-only method for the interaction quench scenario are explained in Sec.~\ref{sec:interaction_quench}. In the beginning of Sec.~\ref{sec:CDW-melting} we also state the new dimensional factorization relation of the CDW order parameter. We close with a summary in Sec.~\ref{sec:summary}.

\section{The Holstein Model}
\label{sec:Holstein-model}

We investigate two quench scenarios in the Holstein model at zero temperature. In both cases the system is taken as half-filled with spin-less electrons on a (hyper-)\linebreak cubic lattice. The respective Hamiltonian is
\begin{equation}\begin{split}
    \hat H &= \omega_0 \sum_i \hat b^\dagger_i \hat b_i - t_h \sum_{\langle i j\rangle}\hat c_i^\dagger \hat c_j - \gamma \sum_i \big( \hat n^{\text{el}}_i - \alpha \big) \big( \hat b_i^\dagger + \hat b_i \big)\\
    &= \hat H_{\text{ph}} + \hat H_{\text{el}} + \hat H_{\text{int}}
    \end{split}
    \label{eq:Holstein_hamiltonian}
\end{equation}
where $\omega_0$ is the energy of the optical phonons (in the Einstein approximation they are taken as dispersionless), $t_h$ is the hopping parameter among neighboring lattice sites and $\gamma$ denotes the electron-phonon coupling strength.
The parameter $\alpha$ allows to either couple the phononic displacement to the electronic density ($\alpha = 0$) or to its fluctuations ($\alpha = \langle n_i^{\rm el} \rangle$).
The creation (annihilation) operator for the phonons and electrons at lattice site $i$ are $\hat b_i^\dagger$ ($\hat b_i$) and $\hat c_i^\dagger$ ($\hat c_i$), respectively. Throughout this paper we use angle brackets to denote ordered pairs (or triples), e.g.~$\langle 1\, 2\rangle \neq \langle 2\, 1\rangle $. Apart from the data shown in Fig.s~\ref{fig:1d_comp} and \ref{fig:ftwa_1d_comp}, where the system has open boundaries, we always assume periodic boundary conditions. The Holstein Hamiltonian couples the local phonon displacement $\hat x^{\text{ph}}_i = (\hat b^\dagger_i + \hat b_i)/\sqrt{2M\omega_0}$ (where $M$ is the oscillator mass) to the electron density $\hat n^{\text{el}}_i = \hat c^\dagger_i \hat c_i$ at the same site, thus modeling the local Coulomb interaction between a charged ion, that can oscillate harmonically around its equilibrium position, and an electron. Though a simplistic approach to describe a polarizable material, this model captures basic screening effects. The underlying microscopic mechanism can also be understood in terms of elastic electron-phonon scattering, which is most easily seen by writing the Hamiltonian \eqref{eq:Holstein_hamiltonian} in reciprocal space.\\
In the remaining part of this section we introduce the two quench protocols.\\

For the quench in the electron hopping $t_h$, we consider $\alpha = 0$ in \eqref{eq:Holstein_hamiltonian} throughout the protocol and start from one of the ground states in the atomic limit (where $\gamma \gg \omega_0 \gg t_h$). This initial state is defined by an alternating occupation of dressed electrons, “polarons”,  on the lattice sites, giving rise to a charge density wave. This can be understood as follows: For $t_h=0$ the Holstein model becomes purely local; $\hat H_{\text{el}}$ is neglected. Then the electron-phonon interaction can be decoupled with a Lang-Firsov transform \cite{Hohenadler2007}\cite{LangFirsov1962}, which is defined by the anti-hermitian operator $\hat S=  -\frac{\gamma}{\omega_0} \sum_l \hat n_l^{\text{el}} ( \hat b^\dagger_l -  \hat b_l) = -\frac{\gamma}{\omega_0}\sqrt{\frac{2}{M\omega_0}} \sum_l \hat n_l^{\text{el}} \hat p^{\text{ph}}_l $, such that
\begin{equation}\begin{split}
    \mathrm e^{\hat S} \left( \hat H_{\text{ph}} + \hat H_{\text{int}} \right) \mathrm e^{\hat S^\dagger} &= \omega_0 \sum_l  \hat b_l^\dagger  \hat b_l - \frac{\gamma^2}{\omega_0} N_{\text{el}} =:  \tilde H_{\text{ph}} + \tilde H_{\text{int}} , \\
    \mathrm e^{\hat S} \hat x^{\text{ph}}_i  \, \mathrm e^{\hat S^\dagger} &= \hat x^{\text{ph}}_i + \frac{\gamma}{\omega_0} \sqrt{\frac{2}{M\omega_0}} \hat n^{\text{el}}_i =: \tilde x^{\text{ph}}_i , \\
    \mathrm e^{\hat S} \hat c_i^\dagger \, \mathrm e^{\hat S^\dagger} &= \mathrm e^{-\frac{\gamma}{\omega_0}( \hat b^\dagger_i -  \hat b_i)}  \hat c_i^\dagger =: \tilde c_i^\dagger ,
    \end{split}
    \label{eq:LF-trf}
\end{equation}
where the tilde denotes a Lang-Firsov transformed quantity, and $N_{\text{el}} = \sum_i \hat n_i^{\text{el}}$. As can be seen from the form of $\hat S$, this transform leaves the phonon momenta invariant and generates a shift of the phonon displacement on lattice sites which are occupied by an electron. Note that for spinful electrons one finds an additional term describing an attractive local interaction of electrons, commonly considered as the origin for Cooper pairing in conventional superconductors.
Searching for the ground states of $\tilde H_{\text{ph}} + \tilde H_{\text{int}}$ restricts the phonon modes to the vacuum $\ket{0}_{\text{ph}}$. No condition on the electronic distribution is imposed so far. For any set $\mathcal L$ containing exactly half of all lattice sites, the state $\ket{\tilde \Psi_0^{\mathcal L}}= \prod_{i\in\mathcal L} \hat c^\dagger_i \ket{0}_{\text{el}}\ket{0}_\text{ph}$ is one valid, degenerate ground state. Inverting the Lang-Firsov transform \eqref{eq:LF-trf}, this state is mapped to
\begin{equation}
    \ket{ \Psi_0^{\mathcal L}}= \prod_{i\in\mathcal L} \mathrm e^{\frac{\gamma}{\omega_0}( \hat b^\dagger_i -  \hat b_i)}  \hat c^\dagger_i \ket{0}_{\text{el}}\ket{0}_\text{ph},
        \label{eq:CDW_initial_state}
\end{equation}
where we can identify the creation operator of a polaron, an electron dressed with a coherent state phonon cloud, $\hat C^\dagger_i := \mathrm e^{\frac{\gamma}{\omega_0}( \hat b^\dagger_i -  \hat b_i)}  \hat c^\dagger_i \neq \tilde c^\dagger_i$. The corresponding ground state energy is $-(\gamma^2/\omega_0) N_{\text{el}}$, which is sometimes termed “polaron binding energy”.

\begin{figure}[tp]
    \begin{minipage}[c]{0.48\linewidth}
            \vspace{4.5em}
            \renewcommand{\arraystretch}{1.6}
            \centering
            \begin{tabular}{c|c} \hline\hline \centering
                 $t_1$ &  $t_h \, \mathrm e^{-\gamma^2/\omega_0^2}$ \\ \hline
                 $V_2$ &  $t_h^2 \omega_0 / \gamma^2$   \\ \hline
                 $t_2$ & $t_h^2 \omega_0 \, \mathrm e^{-\gamma^2/\omega_0^2} / \gamma^2$     \\ \hline\hline
            \end{tabular}
            \captionof{table}{Parameters of the effective Hamiltonian \eqref{eq:H_eff} for $\gamma\gg\omega_0$. }
            \label{tab:eff_params}
            \renewcommand{\arraystretch}{1}
    \end{minipage}%
    \hfill %
    \begin{minipage}[c][5em]{0.48\linewidth}
        \vspace{3em}
        \centering
        (a)
        \includegraphics[width=0.8\linewidth]{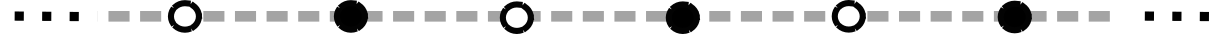}\\\vspace{1em}
        (b)
        \includegraphics[width=0.8\linewidth]{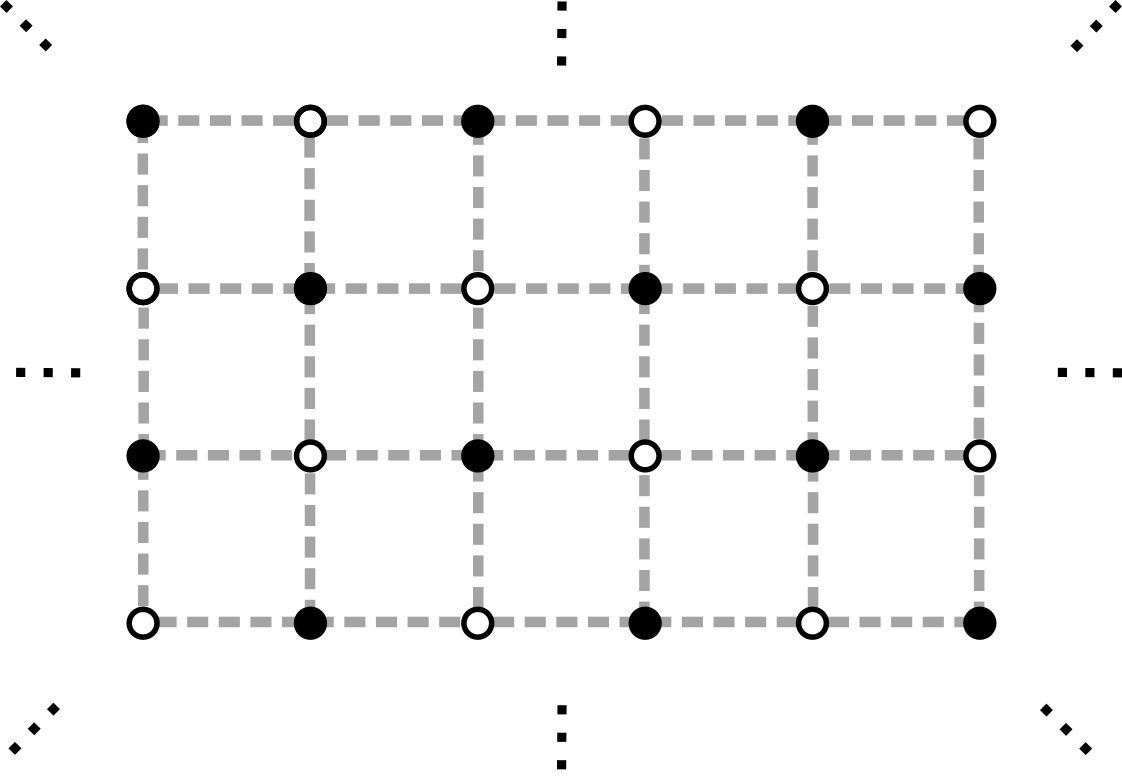}
        \captionof{figure}{The initial electronic occupation for the hopping quench is a charge density wave, here shown for one (a) and two (b) dimensions.}
        \label{fig:checkerboard}
    \end{minipage}
\end{figure}

We can enforce an electronic order by allowing for a small, nonzero hopping $t_h$ to lift the ground state degeneracy. This gives a good approximation especially for small polarons which do not extend beyond one lattice site, hence when the coupling is strong, $2t_h \ll \gamma^2/\omega_0$ \cite{Holstein1959b}. Let P and Q be the projection operators onto the ground state manifold of $\hat H_{\text{ph}} +  \hat H_{\text{int}}$ found above and its orthogonal complement, respectively. We derive an effective Hamiltonian for the ground state manifold by expanding $\hat H$ up to second order in $t_h$ \cite{Mila2010},
\begin{equation}\begin{split}
    &\hat H_{\text{eff}} = \mathrm P  \hat H \mathrm P + \mathrm P \hat H_{\text{el}} \mathrm Q \\
        & \qquad\quad  \cdot \left[-(\gamma^2/\omega_0) N_{\text{el}}  - \mathrm Q \left( \hat H_{\text{ph}} + \hat H_{\text{int}} \right) \mathrm Q \right]^{-1} \mathrm Q \hat H_{\text{el}} \mathrm P.
    \end{split} \label{eq:H_eff_gen}
\end{equation}
\noindent This yields -- up to a constant shift that will be neglected hereafter --
\begin{equation}\begin{split}
        \hat H_{\text{eff}}= & \hat H_{\text{eff,hop}} + \hat H_{\text{eff,nn}} + \hat H_{\text{eff,corr.hop}} \\
            = & -t_1 \sum_{\langle lm\rangle}  \hat C_m^\dagger  \hat C_l + \frac{V_2}{2} \sum_{\langle lm \rangle} \hat n^{\text{el}}_l \hat n^{\text{el}}_m \\                        & \qquad - t_2 \sum_{\stackrel{\langle lmn\rangle,}{l \neq n}}  \hat C_l^\dagger (1-\hat n^{\text{el}}_m)  \hat C_n,
        \end{split}
    \label{eq:H_eff}
\end{equation}
with $\hat n^{\text{el}}_l$ the polaron/electron number at site $l$ and $\langle lmn\rangle$ means that site $m$ is next to sites $l$ and $n$. The values for the non-negative parameters $t_1$, $t_2$ and $V_2$ are summarized in Tab.~\ref{tab:eff_params} for the atomic limit $\gamma \gg \omega_0\,(\gg t_h)$. For the general values and more details on the perturbative expansion with respect to $t_h$ we refer to App.~\ref{app:H_eff}.\\
The nearest neighbor hopping with $t_1$ and the correlated second-nearest neighbor hopping with $t_2$ decay exponentially as $\gamma/\omega_0 \rightarrow \infty$. The repulsive nearest neighbor density-density interaction $V_2$ vanishes only algebraically in the atomic limit and therefore becomes dominating. In this way, an electronic order is established such that for half-filling only every second lattice site is occupied and a charge density wave emerges. For a square lattice this order resembles a checkerboard pattern, see Fig.~\ref{fig:checkerboard}, and similar for higher-dimensional hyper-cubic lattices. The occurence of a CDW state in square lattices is also confirmed by quantum Monte Carlo (QMC) studies.\cite{Costa2020}

For the quench scenario, we assume to start in exactly one of two degenerate ground states in the atomic limit, which differ from each other by a spatial shift of one lattice site or, equivalently, by an inversion of electronic occupation. Let $\mathcal L_{\text{CDW}}$ be the set of lattice sites that are occupied by electrons initially, characterizing the initial state $\ket{\Psi_0^{\mathcal L_{\text{CDW}}}}$, see eq.~\eqref{eq:CDW_initial_state}.
For $t>0$ we will observe a melting of this alternating charge order which can be quantified by an order parameter,
\begin{equation}
    \mathcal O_{\text{CDW}}(t) = \frac{1}{N_{\text{el}}} \sum_l (-1)^{\mathrm{sgn} (l) } \langle \hat n^{\text{el}}_l (t)\rangle,
    \label{eq:O_CDW_def}
\end{equation}
where $\mathrm{sgn} (l) =  0$ if $l\in\mathcal L_{\text{CDW}}$ and 1 else, such that $\mathcal O_{\text{CDW}}=1$ initially. The resulting dynamics will be monitored by the change of the charge density wave order. Complete charge inversion occurs when $\mathcal O_{\text{CDW}}=-1$ and $\mathcal O_{\text{CDW}}=0$ denotes the absence of any charge order of density wave type in the lattice.\\

In addition to the charge density melting, we study a sudden quench of the interaction strength $\gamma$, which allows to shed more light on the methodological challenges for semiclassics in electron-phonon systems.
The initial state is the zero-temperature ground state of the system at $\gamma = 0$, i.e. a product state $\ket{\text{FS}}_\text{el} \ket{0}_\text{ph}$ of the fermionic Fermi sea $\ket{\text{FS}} = \prod_{\epsilon(\vec k) < \epsilon_\text{F}} \hat c_{\vec k}^\dagger \ket{0}_\text{el}$ and the phonon vacuum $\ket{0}_\text{ph}$.
Again, we are interested in the relaxation dynamics after the quench and therefore focus on the electronic distribution function and the various energy contributions from the electronic and phononic subsystems.

\section{Method}
\label{sec:Methods}

In this section we introduce the truncated Wigner formalism for bosons and fermions by explaining its application to the Holstein model \eqref{eq:Holstein_hamiltonian}. For a more in-depth and general guide to TWA we refer to the pedagogical review \cite{Polkovnikov2010}. 
We explain the main concepts of this semiclassical approach for bosonic TWA, while an introduction to fermionic TWA is found at the end of this section.

\subsection{(Bosonic) Truncated Wigner Approximation}
\label{ssec:TWA}

As opposed to the wave function in Schrödinger quantum mechanics, the central object in TWA is the so called Wigner function $W(\psi,\psi^*)$, which depends on the field variables $\psi,\psi^*$, the classical counterpart of the bosonic field operators $\hat\psi,\hat\psi^\dagger$. Also, for any operator $\hat \Omega$, there exists a classical analogue, its so called Weyl symbol $\Omega_W(\psi,\psi^*)$, which is obtained by the Wigner transform of $\hat \Omega$. The Wigner function is simply the Weyl symbol of the density operator. Equivalently, the Weyl symbol can also be computed by symmetrizing the operator w.r.t.~the field operators and then simply substituting $\hat\psi \rightarrow \psi , \hat\psi^\dagger \rightarrow \psi^*$. Any expectation values are then computed by averaging the respective Weyl symbol over all phase space variables, with $W$ acting as pesudo-probability function.
Now, the semiclassical approximation we employ is a truncation of the exact equation of motion of the Wigner function, restricting it to its classical paths and thus making its computation more feasible.
This means, for any time-dependent operator average we evaluate
\begin{equation}
    \langle\hat\Omega(t)\rangle \approx \iint \mathrm d\psi_0 \, \mathrm d \psi_0^* \, W_0(\psi_0,\psi_0^* ) \, \Omega_W (\psi(t),\psi^*(t),t ).
        \label{eq:TWA-avg}
\end{equation}
The field variables $\psi_0,\psi^*_0$ are the initial values for $\psi(t),\psi^*(t)$; both pairs are related via a classical trajectory, which is deterministic and unique. In this way we can interpret the TWA average~\eqref{eq:TWA-avg} as follows: At initial time, the quantum state is described by its Wigner function $W_0$ without any approximations. As time evolves, only classical trajectories of the field variables $\psi,\psi^*$ are considered. Hence, also any Weyl symbols (and their averages) are only evaluated along classical paths in phase space. Quantum fluctuations that deviate from classical trajectories gain significance as time increases such that TWA becomes exact in the short time limit $t\rightarrow 0$.

Practically, the routine goes as follows: one samples $\psi_0, \psi^*_0$ from the initial Wigner function and evolves them classically in time according to the respective classical equations of motion. For any time $t$, one can evaluate the Weyl symbol of the operators of interest, $\Omega_W(\psi(t),\psi^*(t), t)$. Ultimately, one averages over all samples taken to obtain the semiclassical operator average, according to eq.~\eqref{eq:TWA-avg}. Any two semiclassical trajectories are independent from each other. Their computation can be parallelized easily to increase computational efficiency.

The initial sampling of the field variables requires a non-negative initial Wigner function, which is not the case for general quantum states, but holds for our purposes, because the local bosonic states are coherent states- both in the hopping and interaction quench.
Let $b_i,b_i^*$ be the local phononic field variables at lattice site $i$, and $\underline b, \underline b^*$ the corresponding field variable, consisting of the field variables of all sites.
The initial Wigner function for the charge density wave state is the Wigner transform of (the bosonic part of the) the pure state $\ket{\Psi_0^{\mathcal L_{\text{CDW}}}}$,
\begin{equation}\begin{split}
    W^{\text{CDW}}_0(\underline b, \underline b^*) &= \left( \ket{\Psi_0^{\mathcal L_{\text{CDW}}}}\bra{\Psi_0^{\mathcal L_{\text{CDW}}}} \right)_W \\
        &= \prod_{i\in\mathcal L_{\text{CDW}}} 2\,\mathrm e^{-2\big|\frac{\gamma}{\omega_0}-b_i\big|^2} \prod_{j\notin\mathcal L_{\text{CDW}}} 2\,\mathrm e^{-2|b_j|^2} .
    \end{split}
        \label{eq:W_init_CDW}
\end{equation}
In the interaction quench, the phonons are in the vacuum initially, which results in the Wigner function
\begin{equation}
    W^{\text{vac}}_0(\underline b, \underline b^*) = \left( \ket{0}_{\rm ph}\bra{0} \right)_W = \prod_{j} 2\, \mathrm e^{-2|b_j|^2} .
        \label{eq:W_init_int-quench}
\end{equation}
Both Wigner functions are multi-variable Gaussian distributions. In the CDW state, the real part of $\langle \hat b_j \rangle $ (which is proportional to $\langle \hat x^{\rm ph}_j\rangle$) is displaced by $\gamma/\omega_0$ on those lattice sites which host an electron initially.
In order to generate the initial conditions for the classical trajectories, real and imaginary parts of the bosonic variables are independently sampled from the Gaussian distribution function.
If only the phonons are treated semiclassically, the electronic one-particle density matrix is set to its initial state expectation value, which is $\rho_{ij}(0) = \langle c_i^\dagger c_j \rangle_0 = \delta_{ij} n_i$ for a (CDW) product state.

The semiclassical equations of motion for the electronic operators and classical phonon fields in the Heisenberg picture read
\begin{equation}\begin{split}
    -i\hbar \, \frac{\mathrm d{ \hat c_j}}{\mathrm dt} &= \left[ H,\hat c_j \right] = t_h \sum_{i\in \text{NN}(j)} \hat c_i + \gamma\, \hat c_j(b_j + b^*_j), \\
    i\hbar \,\frac{\mathrm d b_j }{ \mathrm dt} &= \frac{\partial H_{ W}}{\partial b_j^*} = \omega_0 b_j -\gamma \left(\langle\hat  n^{\text{el}}_j\rangle -\alpha\right) .
\end{split}
    \label{eq:eom_CDW}
\end{equation}
The nearest neighbor sites of $j$ are denoted by NN($j$) and the substitution $b^{(\dagger)}_j \rightarrow b_j^{(*)}$ has already been inserted in the equation for the $\hat c_j$'s. Because the Holstein Hamiltonian \eqref{eq:Holstein_hamiltonian} conserves the fermionic particle number, we can make the ansatz $\hat c_j(t) = \sum_{j'}A_{jj'}(t) \,\hat c_{j'}(t=0)$ to translate the operator-valued equations to a set of equations for the matrix elements of $A$. In the phononic equations, the electronic number operator has been replaced by its expectation value in an \textit{ad hoc} manner to make the expression complex-valued. There is no \textit{a priori} justification for this, rather crude, approximation. However, the initial state $\ket{\Psi_0^{\mathcal L_{\text{CDW}}}}$ is a \mbox{product} state w.r.t.~the electronic and phononic Hilbert space. Thus, this approximation of mean-field type is expected to preserve the correct short time dynamics of the TWA approach.
The time-dependent one-particle reduced density matrix $\rho_{ij}(t) = \langle \hat c_i^\dagger(t) \hat c_j(t) \rangle$ can be calculated directly from the ansatz and the initial state expectation value:
\begin{align}\begin{split}
\rho_{ij}(t) &= \sum_{m n} A_{i m}^\ast(t) A_{j n}(t) \big\langle \hat c_m^\dagger(0) \hat c_n(0) \big\rangle \\
&= \sum_{m n} A_{i m}^\ast(t) A_{j n}(t) \rho_{m n}(0) .
\end{split}\end{align}
In particular,
\begin{align}\begin{split}
\mathcal O_{\text{CDW}}(t) &= (1/N_{\text{el}})\sum_j \, (-1)^{\text{sgn}(j)} \rho_{jj} \\
&= (1/N_{\text{el}})\sum_j \sum_{j'\in\mathcal L_{\text{CDW}}} (-1)^{\text{sgn}(j)} |A_{jj'}(t)|^2
\end{split}\end{align}
and $n_k(t) :=\langle \hat n^{\rm el}_k(t)\rangle = \frac{1}{\sqrt{V}} \sum_{i} \text{e}^{i \vec k \cdot (\vec r_i - \vec r_j)} \rho_{i j}(t)$. In general, any fermionic $n$-point correlation can be reduced to an expression involving only these time-dependent matrices $A_{jj'}(t)$, if one applies Wick's theorem to the initial fermionic expectation value (that is, rewriting the expectation value at time $t=0$ as an average with respect to the fermionic vacuum $\ket{0}_{\rm el}$). The semiclassical approximation enters fermionic averages only via the electron-phonon coupling to the semiclassical phonon dynamics.\\

\subsection{fTWA+TWA method}
\label{ssec:fTWA}

The TWA idea for bosons has only recently been carried over to fermions~\cite{Davidson2017}, which do not allow for a straightforward semiclassical description.
The fundamental idea of the fermionic TWA (fTWA) is to define Weyl symbols for bilinears $\hat\rho_{ij} = \hat c_i^\dagger \hat c_j - \frac{1}{2} \delta_{ij}$ of the fermionic annihilation and creation operators.
An equivalent method has been developed earlier in the nuclear physics community under the name stochastic mean-field approach~\cite{Ayik2008,Lacroix2014}.
The fTWA method has proven to be useful for disordered systems~\cite{Sajna2020,Iwanek2023}.
In this paper, we propose that applying fTWA and TWA simultaneously to electron-phonon models gives rise to an improvement over the more conventional TWA-only method.
To implement fTWA we use the equations of motion and the same solution strategy with time-dependent coefficients $A_{jj'}(t)$ as outlined above.
Note that the evaluation of expectation values like $\langle \hat c_i^\dagger(t) \hat c_j(t) \rangle$ depends on the initial-time values of $\langle \hat c_i^\dagger \hat c_j \rangle$,
which correspond precisely to the fTWA phase space variables.
Hence, the only change that needs to be done is a sampling of the initial values for the classical $\rho_{ij}$.
The most common strategy is to construct a probability distribution that correctly reproduces means and covariances of a given initial state.
The classical covariance is constructed such that it agrees with the symmetrized quantum correlation of the initial state.
The choice of a distribution function is non-unique and can even influence the predictiveness of the method.
It is in general not possible to construct distributions that correctly reproduce third or higher moments of fermionic many-particle states~\cite{Ulgen2019}.
Like in previous works~\cite{Ulgen2019,Osterkorn2020},
the Gaussian model for the Fermi sea reads as follows (c.c. denotes the connected correlation $\langle a b \rangle^\text{c.c.} = \langle a b \rangle - \langle a \rangle \langle b \rangle$):
\begin{align} \begin{split}
\big\langle \hat \rho_{k l} \big\rangle = \big\langle \hat \rho_{k l}^\dagger \big\rangle &= \delta_{k, l} \left( n_k - \frac{1}{2} \right), \\
 \Big\langle \frac{1}{2} \big\{ \hat \rho_{k l},  \hat \rho_{s p} \big\} \Big\rangle^\text{c.c.} &= \frac{1}{4} \delta_{k p} \delta_{l s} \left( n_k + n_l - 2 n_k n_l \right),
 \label{eq:fermion_wigner_data}
\end{split} \end{align}
where $\{\cdot,\cdot\}$ is the anticommutator for symmetrization.
We also carried out simulations with a discrete “two-point” distribution function $P_{\mu,\sigma}(x) = \frac{1}{2} \big( \delta(x - \mu - \sigma) + \delta(x - \mu + \sigma) \big)$,
which has yielded better results than the Gaussian approximation for certain problems.
In practice, we sample real and imaginary parts independently from the distribution function, both for the phonons (as described in the previous section) and for the fermions according to \eqref{eq:fermion_wigner_data}.
The numerical cost of performing fTWA simulations is the same as performing TWA simulations: A non-linear system of ordinary differential equations needs to be solved for a number of variables that scales quadratically with the system size.

\section{Charge Density Wave Melting}
\label{sec:CDW-melting}

In this section we study the dynamics of the charge density wave after a sudden parameter quench at time $t=0$. We assume that the system has previously been prepared in one of the degenerate ground states of the atomic limit ($\gamma\gg \omega_0 \gg t_h $), which is given by an alternating occupation of polarons on the lattice sites.

\subsection{Factorization of $\mathcal O_{\text{CDW}}$ across dimensions}
\label{ssec:CDW-fact}

\begin{figure}[b!]
    \centering
    \includegraphics[width=\linewidth]{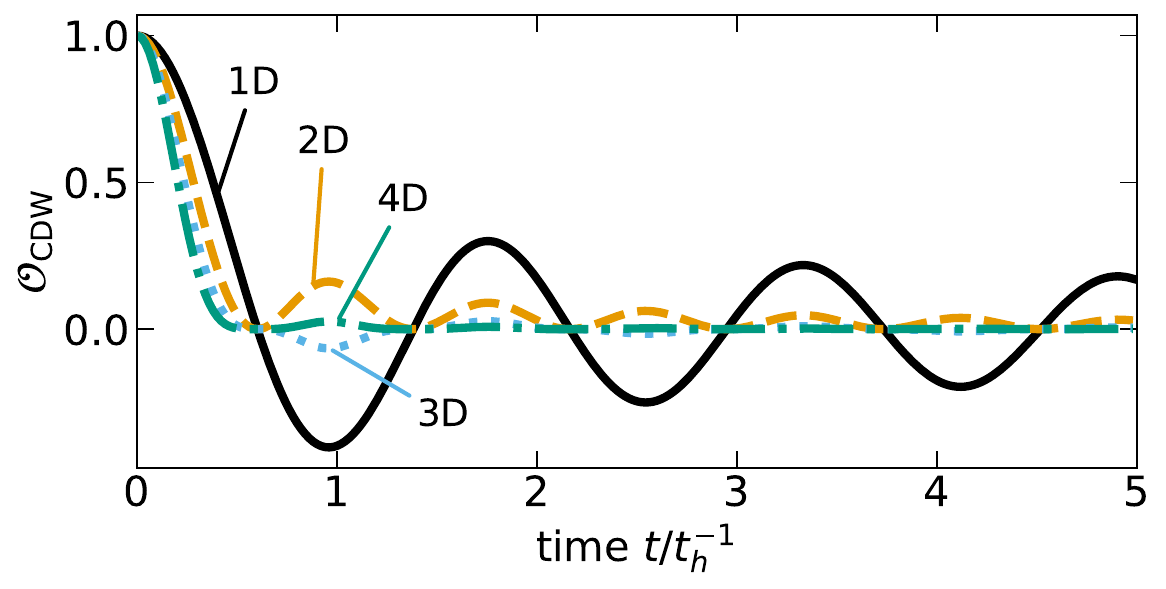}
    \caption{Time evolution of CDW order parameter for \mbox{$d=1,2,3,4$} dimensions on a periodic, hypercubic lattice at $\gamma=0$.}
    \label{fig:ocdw_nocoupling}
\end{figure}%

Starting from a CDW state initially, it can be proven (App.~\ref{app:proof_factorization}) that the order parameter factorizes among the dimensions when the electron-phonon coupling is switched off, i.e.
\begin{equation}
    \left(\mathcal O_{\text{CDW}}^{\gamma=0,1\mathrm{D}}(t)\right)^d = \mathcal O_{\text{CDW}}^{\gamma=0,d\mathrm D}(t) .
    \label{eq:O_cdw_factorization}
\end{equation}
The underlying $d$-dimensional hyper-cubic lattice has periodic boundary conditions.
This is an exact relation and one of the main results of this paper.
The proof of eq.~\eqref{eq:O_cdw_factorization} can be found in Appendix~\ref{app:proof_factorization}.
There, we also prove a generalized factorization relation for primitive triclinic and orthorhombic lattices with periodic boundary conditions. The order parameter in these lattices is isotropic and the relation reads
\begin{equation}
    \mathcal O^{\gamma=0,3\mathrm{D}}_{\mathrm{CDW}}(t) = \mathcal O^{\gamma=0,1\mathrm{D}}_{\mathrm{CDW},\bm i}(t) \cdot \mathcal O^{\gamma=0,1\mathrm{D}}_{\mathrm{CDW},\bm j}(t) \cdot \mathcal O^{\gamma=0,1\mathrm{D}}_{\mathrm{CDW},\bm l}(t),
    \label{eq:O_cdw_fact_gen}
\end{equation}
where $i,j$ and $l$ correspond to the primitive lattive vectors (i.e.~the direction along which the charge difference is taken). Relation \eqref{eq:O_cdw_fact_gen} is exact and holds true as long as next-nearest neighbor hopping and other interaction effects are negligible. For the remaining part of this article though, we focus on the hypercubic lattices.\\
The time evolution of the order parameter at $\gamma=0$ and $d=1,2,3,4$ for a hypercubic lattice is shown in Fig.~\ref{fig:ocdw_nocoupling}. In particular, relation \eqref{eq:O_cdw_factorization} implies that for even dimensions no charge inversion occurs between neighboring lattice sites, as the order parameter will remain nonnegative. Also, with increasing dimension, the oscillations of $\mathcal O_{\text{CDW}}$ are further damped and the characteristic time for the CDW melting reduces. From a microscopic perspective, the coordination number of the lattice regulates the possible electronic hopping processes. In one spatial dimension, CDW order is less likely to break as the Pauli exclusion principle prohibits two electrons on the same lattice site, the overall charge dynamics after the quench follows a more coherent motion than in higher dimensions, i.e.~lattices with larger coordination number.

\subsection{Phonon displacement $x_i^\text{ph}(t)$}
Next to the electronic order parameter $\mathcal{O}_\text{CDW}$,
the expectation value of the phononic position operator $\langle \hat x_i^\text{ph} \rangle = \frac{1}{\sqrt{2\omega_0}} \langle \hat b_i + \hat b_i^\dagger \rangle =: x_i^{\text{ph} }$ is often considered as a complementary order parameter in electron-phonon systems.
It is therefore natural to ask for its time dependence and whether a similar factorization relation can be found.
In an uncoupled system ($\gamma = 0$), the displacement operator obeys the equation of motion $\partial^2_t \hat x_i^\text{ph} = -\omega_0^2 \hat x_i^\text{ph}$ such that, for $\hat p^{\text{ph}}_i(0) = 0$, $ x_i^\text{ph}(t)$ oscillates harmonically between $-x_i^\text{ph}(0)$ and $+x_i^\text{ph}(0)$.
Such dynamics is site-local and, in particular, independent of the dimension.
While $\mathcal{O}_\text{CDW}$ obeys factorization in the uncoupled system, $x_i^\text{ph}$ does not.
In the interacting case, one can derive an exact relation between $x_i^\text{ph}(t)$ and $\mathcal{O}_\text{CDW}(t)$ for the quench problem considered here (details in App.~\ref{app:x_ph}),
\begin{align}\begin{split}
 &x_i^\text{ph}(t) = \frac{\gamma}{\omega_0^{3/2} \sqrt{2}} \Big( 1 \pm \cos(\omega_0 t) \Big) \\
 &\quad \pm \frac{\gamma}{\sqrt{2 \omega_0}} \sin(\omega_0 t) \int_0^t \mathcal{O}_\text{CDW}(\tau) \cos(\omega_0 \tau) \text{d}\tau \\
 &\quad \mp \frac{\gamma}{\sqrt{2 \omega_0}} \cos(\omega_0 t) \int_0^t \mathcal{O}_\text{CDW}(\tau) \sin(\omega_0 \tau) \text{d}\tau .
 \label{eq:x_dyn}
\end{split}\end{align}
The upper signs need to be chosen for an initially occupied site $i$ and the lower ones for an initially empty one.
Dimensional factorization of $\mathcal{O}_\text{CDW}$ clearly does not imply factorization of $x_i^\text{ph}$.\\
In the limit of small changes of the electronic order parameter, $\mathcal{O}_\text{CDW}(t) = 1 - \epsilon(t)$ for $0\leq \epsilon(t)  \ll 1$, and for $x_i^\text{ph}$ being an initially unoccupied site, the following holds: If $\mathcal{O}_\text{CDW}$ obeys the dimensional factorization then
$x_i^{\text{ph,}d\mathrm{D}}(t) = d \cdot x_i^\text{ph,1D}(t) + \mathcal O(\epsilon^2)$. \\
However, we did not find more general relations and will therefore focus on the electronic order parameter in the following.
We want to test its factorization relation beyond the unphysical limit of vanishing electron-phonon interaction $\gamma$, by means of the TWA and fTWA+TWA approaches. Before tackling this question, we benchmark both methods against full-quantum calculations on a Holstein chain.

\subsection{Benchmark results in 1D}
\label{ssec:CDW-1D}

\begin{figure}[b!]
    \centering
    \includegraphics[width=\linewidth]{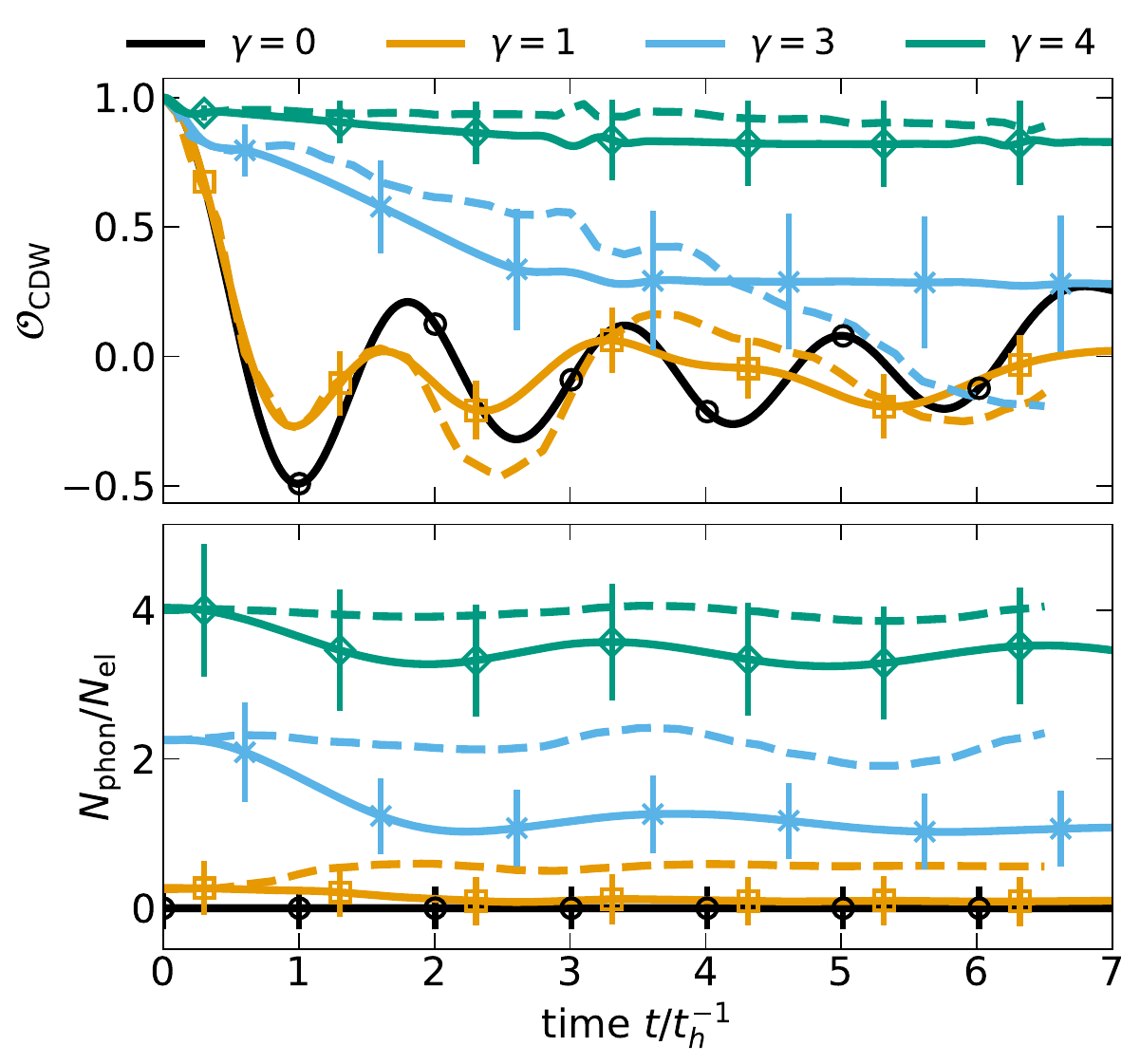}
    \caption{CDW order parameter (top) and phonon number per electrons (bottom) as a function of time after the quench to $\omega_0 / t_h= 2 $ and $\gamma /t_h= 0,1,3,4$. Solid lines give TWA, dashed lines tDMRG \cite{Stolpp2020} data. The open Holstein chain has a length of $L=13$ and $N_{\text{el}}= 6$. Shown data was averaged over $3\cdot 2^9$ sample trajectories.}
    \label{fig:1d_comp}
\end{figure} %
\begin{figure}[t!]
    \centering
    \includegraphics[width=\linewidth]{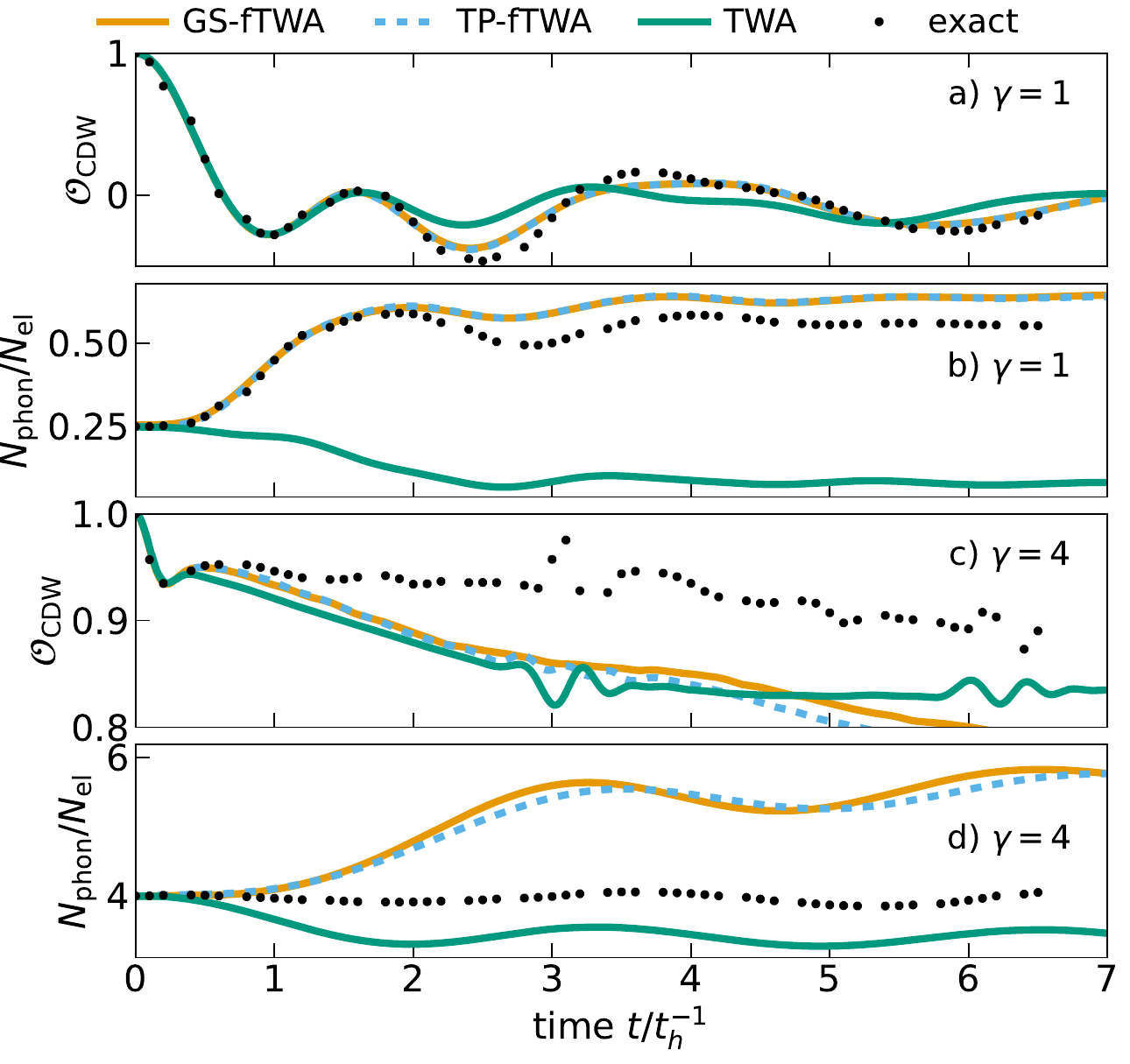}
    \caption{Charge density wave melting on a chain with length $L = 13$ and open boundary conditions.
    $\omega_0/t_h = 2$. Method comparison of fTWA+TWA with a Gaussian Wigner function model (GS-fTWA), a twopoint distribution (TP-fTWA) and TWA without fTWA.
    The data is averaged over 25600 trajectories. Black dots: tDMRG data by Stolpp et al.~\cite{Stolpp2020}
    \label{fig:ftwa_1d_comp}}
\end{figure}%
To investigate the validity of our semiclassical approach, we compare the charge density wave melting with results obtained by full-quantum time-dependent density-matrix renormalization group (tDMRG) calculations for an open Holstein chain \cite{Stolpp2020}. Fig.s~\ref{fig:1d_comp} and \ref{fig:ftwa_1d_comp} show the dynamics of the CDW order parameter and the phonon number at $\omega_0/t_h = 2$ and varying coupling strength $\gamma$ for the TWA and fTWA+TWA method, respectively. In Fig.~\ref{fig:1d_comp}, the errorbars at every 50th data point illustrate the stochastic nature of the semiclassical time evolution, even when the data has converged w.r.t.~samplesize. We omit these in any following figures in favor of clarity.\\
We observe charge density wave melting for all chosen coupling strengths with a characteristic time that increases with the coupling. The effective strong-coupling Hamiltonian \eqref{eq:H_eff} for $\gamma\gg \omega_0\gg t_h$ predicts the phase transition point between a CDW phase and Luttinger liquid at $\gamma/t_h \approx 2.6$ \cite{HirschFradkin1983}. However, DMRG \cite{Bursill1998} and QMC \cite{Creffield2005} studies show this is a slight underestimate for the actual transition point, such that we can consider the data at $\gamma=4$ to lie within the CDW phase. 
In this case, the CDW melts only very slowly. The phonon number, initially at $(L-1)\gamma^2/(2\omega_0^2)$, remains largest throughout the time evolution, even though (fTWA+)TWA systematically \mbox{(over-)}underestimates its value at later times, see Fig.~\ref{fig:ftwa_1d_comp}.
As $\gamma/\omega_0$ is increased,
the initial, local Gaussian Wigner functions (see eq.~\ref{eq:W_init_CDW}) on lattice sites with electrons become peak-like with diverging mean; their constant width, caused by quantum fluctuations, becomes negligible.
Therefore, this regime should be particularly suited for a semiclassical description.
\begin{figure*}[t!]
    \centering
    \includegraphics[width=\textwidth]{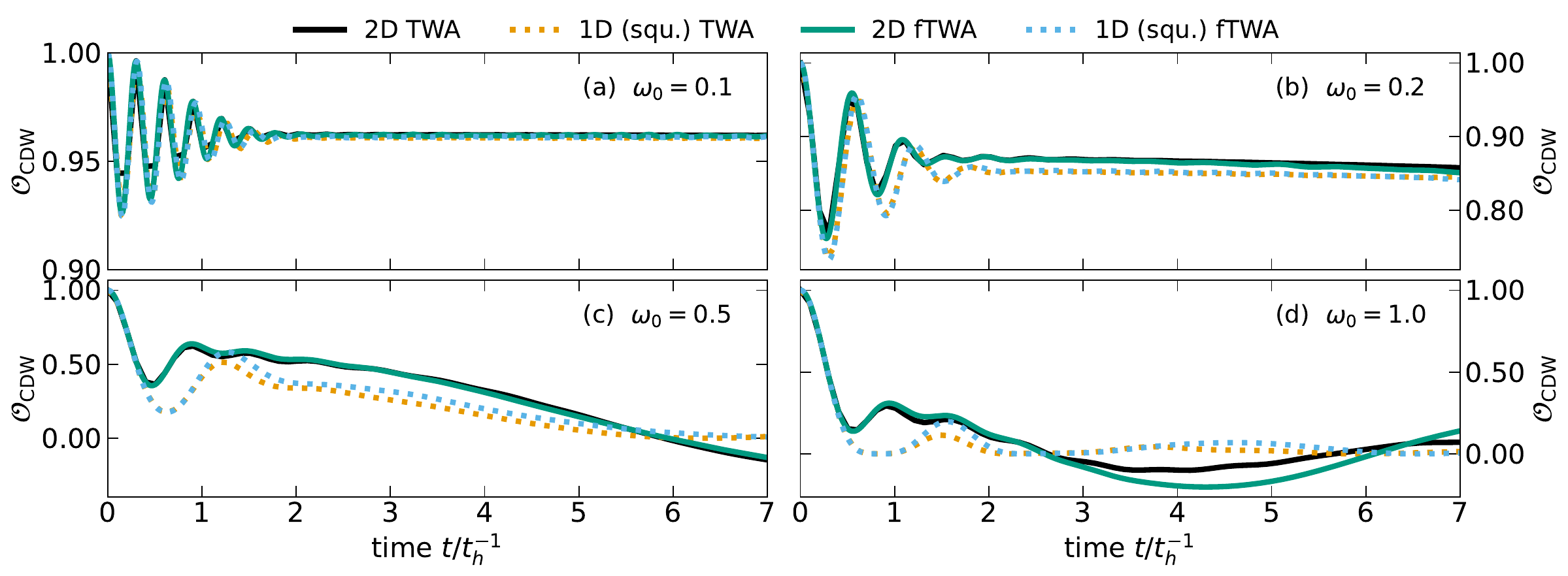}
    \caption{CDW order parameter on a periodic, 20x20 square lattice (2D, solid lines) and the squared CDW order parameter for a periodic chain (1D, dotted) with 30 sites, as function of time at $\gamma/t_h= 1$, $\omega_0/t_h = 0.1, 0.2, 0.5, 1.0$ ((a) - (d)). Black and orange (teal and blue) lines show (Gaussian fTWA+)TWA data, averaged over $3\cdot 2^9$ (TWA) and $3200$ (fTWA+TWA) sample trajectories. }
    \label{fig:ocdw_weakcoupling}
\end{figure*}
For $\gamma\leq 1$, partial charge inversion occurs as the order parameter oscillates around zero within a few hopping times $t_h^{-1}$. 
We verified that for $\gamma=0$, the shown results match the exact, analytical formula \eqref{eq:O_CDW_1D}; the TWA results become exact for free particles without scattering and any stochastic errorbars vanish. The overshoot at $t\gtrsim 6.5 t_h$ stems from self-interference effects due to the finite size of the chain.
For all coupling strengths $\gamma$, Fig.~\ref{fig:1d_comp} illustrates quantitative agreement of TWA and DMRG data for the CDW order parameter at early times, whereas details in the later dynamics are not captured by the semiclassical approach and only a common qualitative behavior remains.
This observation is consistent with the discussion in Section \ref{sec:Methods}, as TWA is exact for the initial time but neglects quantum fluctuations from the path integral that become increasingly important with proceeding time.
In contrast to the CDW order parameter, the TWA result for the phonon number is not correct at early times.
While the tDMRG data shows an initial increase of $N_\text{phon}$, the TWA data shows a decrease there. This is consistent with an analytical short-time expansion of the respective equations of motion and can be traced back to a neglect of fermion correlations of the initial state. We discuss the issue in more detail for the simpler interaction quench problem in Sec.~\ref{sec:interaction_quench}.
The problem can be resolved by extending the TWA to the fTWA+TWA method, see Fig.~\ref{fig:ftwa_1d_comp} for $\gamma/t_h=1,4$.
At $\gamma/t_h = 1$, the agreement of the semiclassical order parameter with the exact dynamics is even significantly enhanced. 
Gaussian and two-point distribution function for the fTWA sampling yield similar results for the CDW order parameter.
For $\gamma/t_h = 4$, the additional fTWA sampling yields small improvements at early times but no systematic improvement at later times. The two-point distribution function seems to perform slightly better than the Gaussian.
In particular, the weak order parameter oscillation around time $t = 3$, which is visible both in the exact and in the TWA data, is only reproduced (with a smaller amplitude than for the exact data) for the two-point distribution. \\
To summarize, we have found a qualitative description of the CDW melting within the (fTWA)+TWA that agrees with the exact tDMRG results for the parameter sets shown.
TWA predicts the short-time dynamics of the order parameter correctly but only fTWA+TWA reproduces the exact phonon number dynamics.
As discussed above, for weak coupling $\gamma\rightarrow 0$ and strong coupling $\gamma/\omega_0 \rightarrow \infty$, the accuracy of our semiclassical approaches is best.
In the following, we extend the study to the two-dimensional Holstein model on a cubic lattice, focusing on the new order parameter factorization \eqref{eq:O_cdw_factorization} in the weak and strong coupling regime.

\subsection{Factorization of $\mathcal O_{\text{CDW}}$ in 2D}
\label{ssec:CDW-nD}

\paragraph*{Numerical results ---} With our semiclassical approaches we are now able to test the factorization relation \eqref{eq:O_cdw_factorization} of the order parameter \textit{beyond} the no-coupling limit, for one and two dimensions. We study its validity in the weak and strong coupling regime, where (fTWA+)TWA give most accurate results. In both cases, we restrict to $\gamma/\omega_0 \gg 1$ to ensure a large number of phonons initially, which justifies the TWA description. 

For weak coupling ($\omega_0 \ll \gamma \lesssim t_h$) Fig.~\ref{fig:ocdw_weakcoupling} explores the similarity between $\left(\mathcal O_{\text{CDW}}^{1\mathrm D}(t)\right)^2$ and $\mathcal O_{\text{CDW}}^{2\mathrm D}(t)$. Their initial and short time overlap, that can be observed for all parameter sets shown, drops off at larger times. The divergence sets in after a time that depends on $\gamma/\omega_0$; the deeper we quench in the CDW phase ($\gamma/\omega_0$ larger), the longer the factorization relation \eqref{eq:O_cdw_factorization} remains valid.

\begin{figure}[t!]
    \centering
    \includegraphics[width=\linewidth]{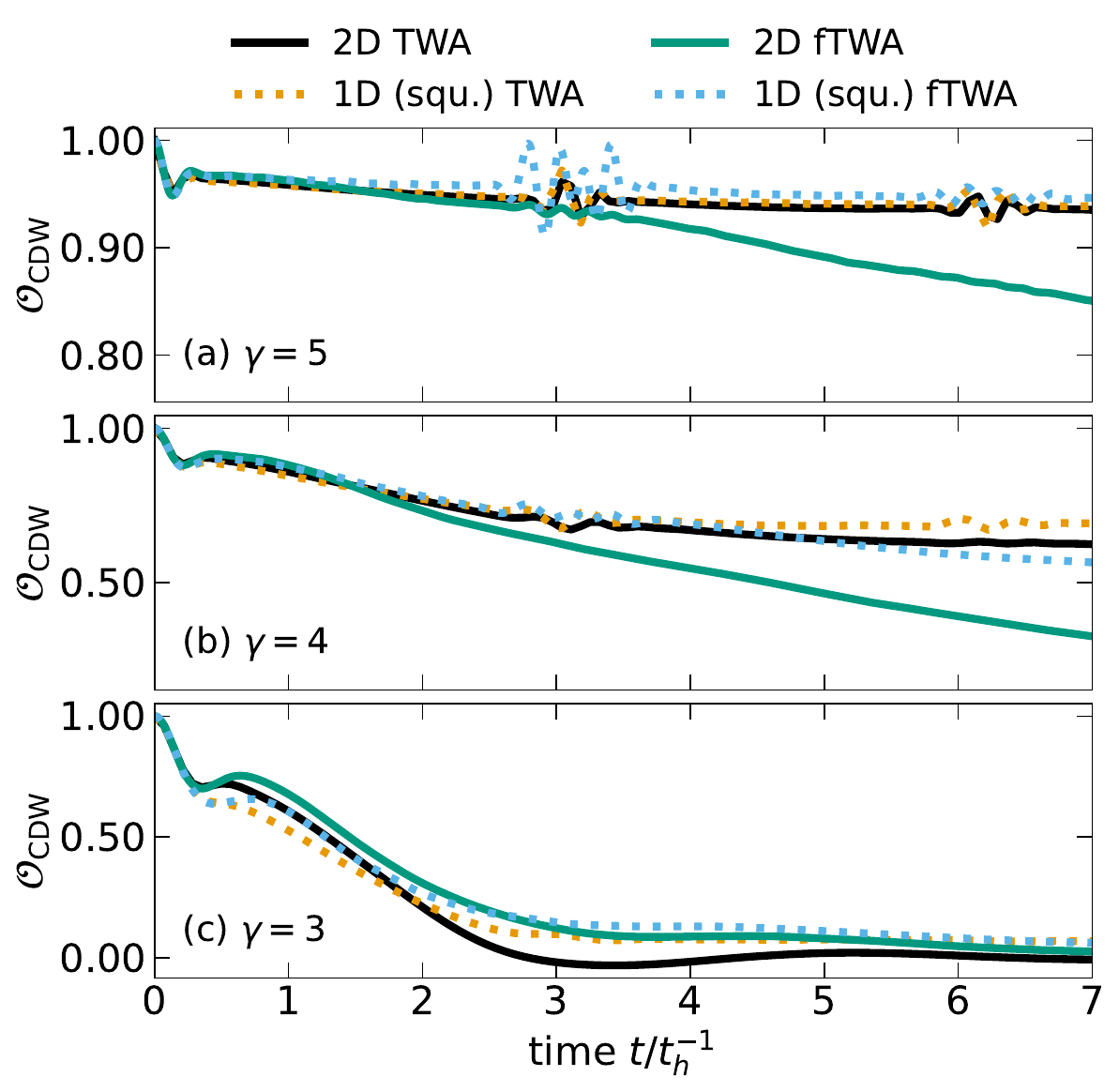}
    \caption{CDW order parameter on a periodic, 20x20 square lattice (2D, solid lines) and the squared CDW order parameter for a periodic chain (1D, dotted) with 30 sites, as function of time at $\omega_0/t_h= 2$, $\gamma /t_h = 5,4,3$ ((a) - (c)). Black and orange (teal and blue) lines show (two-point fTWA+)TWA data, averaged over $3\cdot 2^9$ (TWA) and at least $1600$ (fTWA+TWA) trajectories. }
    \label{fig:ocdw_strongcoupling}
\end{figure}

In the strong-coupling regime ($\gamma\gg \omega_0 > t_h$), polarons form and their dynamics is governed by the effective Hamiltonian \eqref{eq:H_eff} deep in the CDW phase.
For a CDW initial state, only the renormalized hopping ($\propto t_1$) contributes to the short time dynamics, which makes the system an approximate free polaron gas.
Hence, we also expect the order parameter to factorize in this regime. Fig.~\ref{fig:ocdw_strongcoupling} confirms this for two dimensions.
The agreement of the squared order parameter in one dimension and the order parameter in two dimensions can be seen for the TWA and fTWA+TWA especially in the initial quench dynamics, while an increase in the phonon number ($\gamma/\omega_0$ larger) even further improves this correspondence, just as in the weak-coupling case (note the different scale of the axes in the panels in Fig.~\ref{fig:ocdw_strongcoupling}). However, also for later times the factorization relation seems to hold on a qualitative level, although fTWA+TWA predict a faster melting in 2D for $\gamma=4$ \& $5$ after the short time regime than only phononic TWA.
In Fig.~\ref{fig:ocdw_strongcoupling}, we show fTWA data obtained from the sampling of the two-point distribution function since it reproduces the small order parameter oscillations occuring at strong coupling better than the Gaussian distribution.

\paragraph*{Analytical estimate ---}From a short-time expansion of $\mathcal O_{\text{CDW}}(t)$ (around \mbox{$t=0$}) with respect to the corresponding effective Hamiltonian $\hat H_{\text{eff}}$ \eqref{eq:H_eff}, we can estimate the time after which the free polaron approximation is violated (see App.~\ref{app:H_eff} for details). The first term deviating from free polaron dynamics, that is, time evolution with respect to $\hat H_{\rm eff,hop}$, is cubic in time,
\begin{equation} \begin{split}
    &\mathcal O_{\rm CDW}(\delta t) = \,\mathcal O_{\rm CDW}^{\rm eff,hop}(\delta t) +i\, \delta t^3 \\
    & \qquad \times \left\langle \hat H_{\rm eff,hop} \left[ \hat n_j^{\rm el}, \hat H_{\rm eff,corr.hop} + \frac{\hat H_{\rm eff,nn}}{3} \right] \hat H_{\rm eff,hop} \right\rangle,
\end{split} \label{eq:factorization_corr}
\end{equation}
with $j\in \mathcal L_{\text{CDW}}$ and the expectation value is with respect to the initial CDW state $\ket{\Psi_0^{\mathcal L_{\text{CDW}}}}$. Thus, we expect to observe a factorization of the order parameter at least up to second order in $\delta t$, however, with a renormalized characteristic oscillation period compared to the case $\gamma = 0$, cf.~Fig.~\ref{fig:ocdw_nocoupling}, $t_h\rightarrow t_1$. Physically, the faster melting of the CDW with increasing dimension, as also seen in Fig.~\ref{fig:ocdw_nocoupling}, can be attributed to the higher number of possible hopping and correlated hopping processes in higher-dimensional hyper-cubic lattices. Free motion of the polarons is broken by the terms \eqref{eq:factorization_corr} which, in particular, allow correlated hopping processes. They, in turn, are restricted or enabled by the geometry, and hence, the dimension of the hyper-cubic lattice.
After multiplying out the effective parameters $t_1, t_2, V_2$, the remaining non-zero scalar products in \eqref{eq:factorization_corr} are of order unity. Therefore, we estimate the factorization relation to hold for times 
\begin{equation}
    t \ll \text{min}\left( (t_1^2 t_2)^{-1/3} , (3/(t_1^2 V_2))^{1/3} \right).
    \label{eq:temp_estimate_factorization}
\end{equation}
For the regime $\gamma/\omega_0 \gg 1$, this estimate amounts to $t \ll \omega_0 \sqrt[3]{3 \exp(2\gamma^2/\omega_0^2) / (4t_h^4 \gamma^2)}$.
Evaluating eq.~\ref{eq:temp_estimate_factorization} for $\omega_0/t_h = 2$ and $\gamma/t_h = 3, 4, 5$, as shown in Fig.~\ref{fig:ocdw_strongcoupling}, we find $9.8 \,t_h^{-1}, 39\, t_h^{-1}, 209\, t_h^{-1}$ resp.~(the exact values for $t_1, V_2$ and $t_2$ are found in Tab.~\ref{tab:eff_params_gen}). All of these bounds exceed the validity range of the (fTWA+)TWA, which we have discussed in the previous section \ref{ssec:CDW-1D}, thus being even beyond the short time dynamics our semiclassical methods can describe. Hence, the difference of the squared one-dimensional and the two-dimensional order parameter, obtained by the fTWA+TWA method, in Fig.~\ref{fig:ocdw_strongcoupling}(a) \& (b), most likely results from a breakdown of the semiclassical description and not from the violation of the factorization relation -- which we expect to happen at much later times.

\section{Interaction quench}
\label{sec:interaction_quench}

The charge density wave melting problem demonstrates that TWA is able to capture reasonably well the dynamics of the CDW order parameter but fails to describe the phonon number dynamics already at early times.
To shed more light on this aspect, we consider a complementary problem, namely a quench of the electron-phonon coupling parameter.
This problem was studied by Murakami et al.~\cite{Murakami2015} using non-equilibrium dynamical mean-field theory (DMFT).
Following their example, we set $\alpha = \langle \hat n_i^{\rm el}(t=0) \rangle = \frac{1}{2}$ in \eqref{eq:Holstein_hamiltonian},
which eliminates the mean-field phonon dynamics from the problem and allows to focus on correlation-induced effects.
The system is initially prepared in the ground state of the uncoupled electron-phonon system ($\gamma = 0$), which corresponds to the Fermi sea in the electronic and to the vacuum in the phononic part, respectively.

At time $t=0$, the phonon occupation number is zero. After switching on the interaction, we expect energy to be injected into the phononic sector and therefore the phonon number should increase with time.
In order to develop some analytical expectations of the early-time dynamics of the phonon numbers,
we carried out a short-time expansion of the Heisenberg equation of motions:
\begin{align}\begin{split}
\label{eq:Number_Operator_Expansion}
 \langle \hat b_i^\dagger \hat b_i \rangle(t) &= \langle \hat b_i^\dagger \hat b_i \rangle(0) + \big\langle \partial_t \big( \hat b_i^\dagger \hat b_i \big) \big\rangle(0) \cdot t \\
 &\quad + \frac{1}{2} \big\langle \partial_t^2 \big( \hat b_i^\dagger \hat b_i \big) \big\rangle(0) \cdot t^2 + \dots
\end{split}\end{align}
Straightforward calculation 
yields
\begin{align}\begin{split}
\label{eq:Number_Operator_Coefficients}
\partial_t \big( \hat b_i^\dagger \hat b_i \big) &= \gamma \sqrt{\frac{2}{\omega_0}} \big( \hat n^{\rm el}_i - \alpha \big) \,\hat p^{\rm ph}_i \\
\partial_t^2 \big( \hat b_i^\dagger \hat b_i \big) &= \frac{\gamma t_h}{\sqrt{\omega_0}} \sum_{l \in \text{NN}(i)} \hat j_{i,l} \, \hat p^{\rm ph}_i + 2 \gamma^2 \big( \hat n^{\rm el}_i(0) - \alpha \big)^2 \\
&\quad - \sqrt{2} \gamma \omega_0^{3/2} \big( \hat n^{\rm el}_i - \alpha \big) \hat  x^{\rm ph}_i ,
\end{split}\end{align}
where $\hat j_{i,l} = i \big(\hat c_i^\dagger \hat c_l - \hat c_l^\dagger \hat c_i \big)$.
Since phonon vacuum expectation values vanish for terms linear in the $\hat x^{\rm ph}_i$ or $\hat p^{\rm ph}_i$ operators, only the term
\begin{align}\begin{split}
\langle \hat b_i^\dagger \hat b_i \rangle(t) &= \gamma^2 t^2 \big\langle \big( \hat n^{\rm el}_i(0) - \alpha \big)^2 \big\rangle \\
&= \gamma^2 t^2  \big\langle \hat{n}^{\rm el}_i(0)^2 + \alpha^2 - 2 \alpha \hat n^{\rm el}_i(0) \big\rangle
\end{split}\end{align}
is non-zero for both the quantum and the Wigner function expectation value.
Making use of the fact that for (spinless) fermions $\hat{n}^2 = \hat{n}$ and $\langle \hat n^{\rm el}_i(0) \rangle = \alpha = \frac{1}{2}$, we get
\begin{align}\begin{split}
\langle \hat b_i^\dagger \hat b_i \rangle(t) &= \gamma^2 t^2 \big( \alpha + \alpha^2 - 2 \alpha^2 \big) \\
&= \gamma^2 t^2 \big( \alpha - \alpha^2 \big) = \frac{\gamma^2 t^2}{4}
\label{eq:int_quench_short_time}
\end{split}\end{align}
In case of the bosonic TWA, we perform the mean-field substitution for the fermion density $\hat{n} \rightarrow  \langle \hat{n} \rangle$ right in the equations of motion, as a consequence of which the early-time expansion obtained via TWA yields
\begin{align}\begin{split}
\langle b_i^\ast b_i \rangle(t) &= \gamma^2 t^2 \big( \langle \hat{n}_i(0) \rangle - \alpha \big)^2 = 0 .
\end{split}\end{align}
Further inspection of the equations of motion shows that the first non-vanishing term in the TWA short-time expansion is quartic with \textit{negative} prefactor.
These calculations confirm explicitly that plain TWA does not correctly describe the phonon number dynamics,
which can be traced back to an insufficient representation of fermion correlations of the initial state,
despite the usage of the quantum equations of motion for the fermions.
The exact short-time dynamics of the CDW melting problem has a vanishing $\sim t^2$-term because in this case the second and third terms in \eqref{eq:Number_Operator_Coefficients} cancel exactly (note that there $\alpha = 0$ and $\langle \hat{x}_i^\text{ph} \rangle = \sqrt{2} \gamma / \omega_0^{3/2}$ for initially occupied sites).
Hence, one would need to push the short-time expansion to higher orders to see the origin of the TWA failure there.

\begin{figure}[t!]
 \centering
 \includegraphics[width=\linewidth]{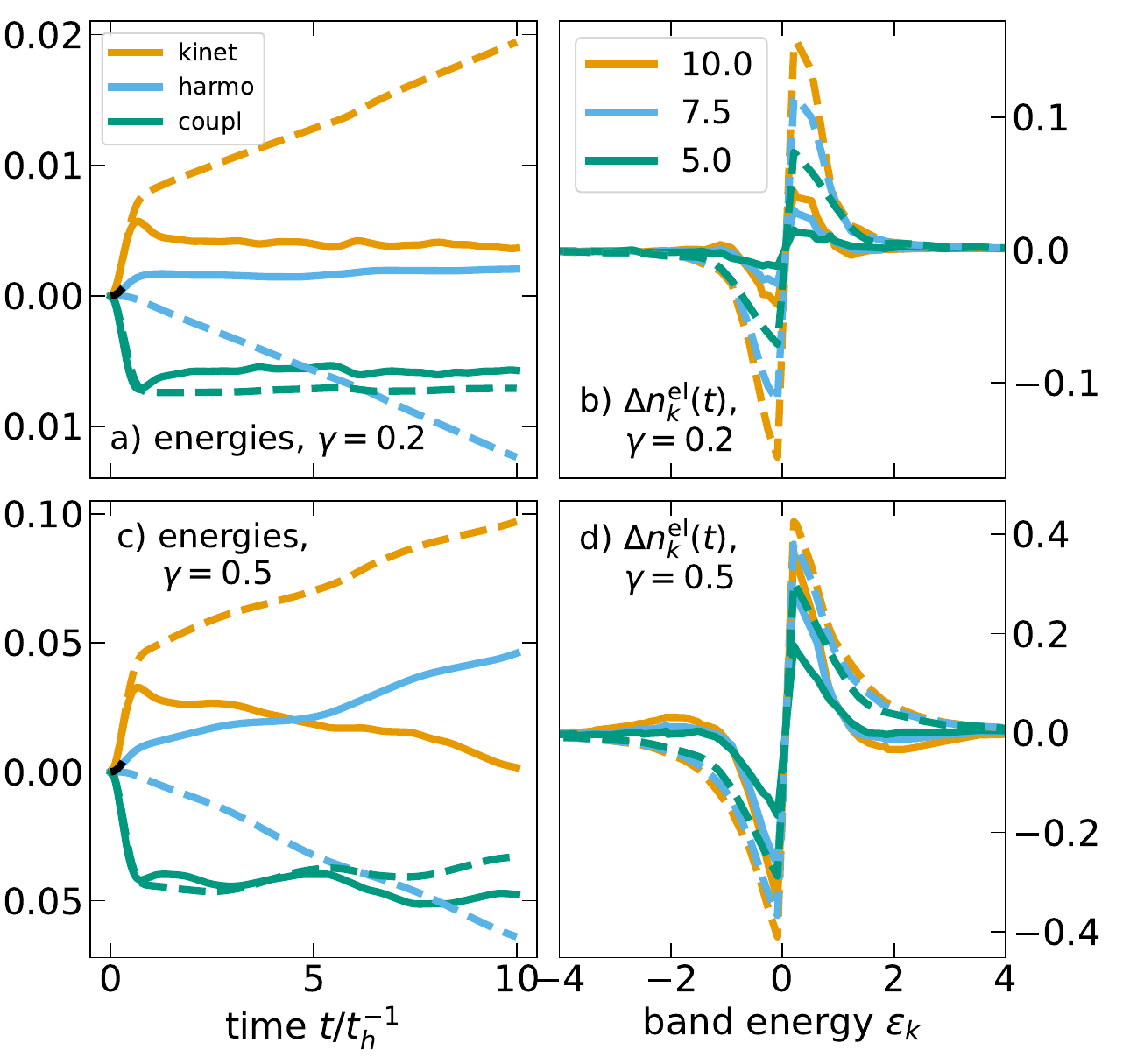}
 \caption{Numerical results for the interaction quench in a 3D cubic lattice with $11^3$ sites and for $\gamma/t_h = 0.2$ and $\gamma/t_h = 0.5$, respectively. Solid lines: Gaussian fTWA+TWA (1000 trajectories), dashed lines: TWA (1000 trajectories). Panels a) and c): change of energies $\Delta E = E(t) - E(0)$, b) and d) electron distribution function $\Delta n_k(t) = n_k(t) - n_k(0)$ at different times (for better visibility, the $x$-axis is limited to $[-4,4]$). Black line: Short-time dynamics result~\eqref{eq:int_quench_short_time}}
 \label{fig:energies_quench-gamma_3d_L11}
\end{figure}

The fTWA method~\cite{Davidson2017} reintroduces fermion correlations to some extent via a suitably constructed Wigner function for the bilinear expectation values $\langle \hat c_i^\dagger \hat c_j \rangle$.
Fig.~\ref{fig:energies_quench-gamma_3d_L11} shows a comparison of TWA and fTWA+TWA numerical data for the interaction quench problem in a three-dimensional lattice with $11^3$ sites and periodic boundary conditions.
After the sudden quench the kinetic energy rises while the electron-phonon coupling energy shrinks.
The early time behaviors of these two energy contributions is similar for the two methods.
A significant deviation can be seen in the behavior of the phonon energy:
In agreement with~\eqref{eq:int_quench_short_time}, it grows quadratically for fTWA+TWA but becomes negative immediately within TWA.
For both methods, the increase of kinetic energy is accompanied by a reduction of the jump at the electron distribution function at the Fermi energy.
In Fig.~\ref{fig:energies_quench-gamma_3d_L11}b) and d) we plot the change $\Delta n_k(t) = n_k(t) - n_k(0)$ for quenches to $\gamma/t_h = 0.2$ and $\gamma/t_h = 0.5$, respectively.
The decrease of the Fermi surface discontinuity is faster for TWA than for fTWA+TWA, which is consistent with the stronger growth of kinetic energy.
Nevertheless, this increase in the TWA case is significantly influenced by the conservation of the total energy and the unphysical dynamics of the phonon energy contribution.
Therefore,
the late-time TWA dynamics cannot be trusted in this case.
Another interesting observation can be made when looking at the fTWA+TWA results for $\Delta n_k(t)$.
At late times, e.g. $t = 10$ for $\gamma/t_h = 0.5$, the distribution function develops values outside of the physical interval $[0,1]$.
This behavior of the fTWA method is also known from previous applications and indicates the end of its range of validity.
The overall fTWA+TWA dynamics yields a similar picture to the one presented in Murakami et al.~\cite{Murakami2015}:
after a rapid initial change of the energies,
the system reaches a quasi-stationary state. Their Bethe lattice DMFT study shows an additional decay of initial energy oscillations, which are not present in Fig.~\ref{fig:energies_quench-gamma_3d_L11}.
This is likely due to the different lattice geometry and potentially a different parameter regime.
To get an intuition for the failure of TWA in the case of the interaction quench problem, we point out that for the phonon dynamics we make a substitution of the electron density operator by its expectation value.
For the CDW case, the fluctuations of $\hat{n}_i^\text{el}$ in the initial state are zero, while they are maximal ($ = \frac{1}{4}$) in the case of the Fermi sea.
This explains that the mean-field replacement is a particularly bad approximation to the interaction quench problem and not so bad for the CDW melting.

\section{Conclusions}
\label{sec:summary}
Nonequilibrium dynamics in charge density wave systems is a challenge for theoretical modeling due to the coupled electronic and lattice degrees of freedom.
While two- and three-dimensional systems are of particular interest experimentally, a lack of reliable computational methods prohibits simulations over long times.
As a prototypical problem we studied the melting of charge density waves in Holstein models in two dimensions.
In order to be able to compare to exact data in one-dimensional systems, we concentrated on the simple protocol of a sudden switch-on of the hopping between lattice sites. 
We approached the dynamics from two perspectives:
In the non-interacting limit, the CDW order parameter on a $d$-dimensional hypercubic lattice can be written as the order parameter on a chain, raised to the power $d$, for all times.
We argued that in the regime of strong electron-phonon coupling $\gamma \gg \omega_0$, the CDW order parameter in high-dimensional Holstein models also satisfies this relation up to times exponential in $(\gamma/\omega_0)^2$.
This order parameter factorization holds true for finite systems as well as in the thermodynamic limit. Additionally, we also proved a generalized factorization relation in the non-interacting limit for primitive triclinic and orthorhombic lattices with periodic boundary conditions. Also, all implications for the factorization in the strong coupling regime should follow in the same way as for hypercubic lattices.
The quench dynamics of the order parameter in other bipartite, non-Bravais lattices and whether a dimensional relation can be found there, might be an interesting topic for future studies.\\
Our second, numerical, approach is based on truncated Wigner approximations.
We compare phonon-only TWA, a method which is also known under the name  “multi-trajectory Ehrenfest approach”\cite{tenBrink2022}, to a semiclassical treatment of both electrons and phonons (“fTWA+TWA”).
Both methods yield results consistent with the factorization property in the regime $\gamma \gg \omega_0$.
Furthermore, they indicate that its validity even extends beyond strong coupling.
Connecting the dynamics of order parameters in higher-dimensional lattices to lower-dimensional ones is a way to develop approximations in otherwise numerically inaccessible regimes.
Our study motivates a further exploration of models and parameter regimes, for which such approximations work.
Next steps could include more complicated excitation protocols like optical driving instead of a quench or spatially inhomogeneous setups~\cite{Jansen2021}.
Beyond the strong coupling limit of the Holstein model, where the semiclassical description is justified by the presence of phononic coherent states, one could also make use of the semiclassical, adiabatic regime $\omega_0 \ll t_h$ when investigating more complex nonequilibrium protocols.
The close relation between the one-dimensional nearest-neighbor density-density interacting model (the strong coupling limit of the Holstein model; eq.~\eqref{eq:H_eff} for $\gamma/\omega_0 \gg 1$, i.e.~$t_2=0$) and the XXZ spin chain suggests further research in the direction of spin models and magnetization dynamics~\cite{Barmettler2009}\cite{HirschFradkin1983}.
The Jordan-Wigner transformation, which maps between both models, translates the CDW order parameter to the staggered magnetization. 
Thus, it seems natural to ask whether a similar factorization relation of the staggered magnetization exists when quenching from the Néel state of the antiferromagnetic phase.

A comparison with exact time-evolution data for a Holstein chain reveals that the additional TWA treatment of the fermions improves the agreement with the reference data, especially in the weak-coupling regime.
In particular, without the additional semiclassical treatment of the electrons, the phonon number does not agree with the exact data already at early times.
To highlight the necessity of sampling both the phononic and electronic initial states, we also investigated a quench in the electron-phonon coupling strength (starting from the Fermi sea and phonon vacuum) as a second application example.
In this case, the phonon-only TWA yields negative phonon occupation numbers for all times.
In contrast, fTWA+TWA  yields the correct early-time dynamics, but tends to develop negative electron occupation numbers at late times.
This clearly indicates the end of the range of validity of the method.
We have, nevertheless, demonstrated how to improve an established semiclassical electron-phonon method.
Despite these intrinsic limitations of the approximations involved, the numerical solution of the method can, however, be pushed to several hundreds of time units (e.g. inverse hoppings) since only mean-field equations of motion need to be solved.
Typically, the later the time, the more trajectories need to be taken into account in order to reach convergence of the ensemble averages.
Lattices sizes of $10 \times 10$, for instance, can still be simulated on a laptop.
Further development should focus on the question how to prevent the fTWA dynamics from developing unphysical electron occupation numbers at late times.
One possible direction could be the enhancement of the phase space by including more variables~\cite{Davidson2015,Czuba2020}.
Finally, we like to point out the fact that semiclassical methods require much less computational resources than Hilbert space methods since the number of dynamical variables scales only quadratically with the number of lattice sites.
The independence of individually sampled trajectories from each other allows for straightforward parallelization.

In addition, electronic susceptibilities and n-point correlation functions are also easy to obtain within the semiclassical description and will be a worthwhile topic of future work.

\section*{Acknowledgments}
We thank Anatoli Polkovnikov, Michael ten Brink, Stefan Gr\"{a}ber and Yuta Murakami for insightful discussion on the topic of this paper.
E.P.~acknowledges funding from the Cluster of Excellence "CUI: Advanced Imaging of Matter" of the Deutsche Forschungsgemeinschaft (DFG)-390715994/EXC 2056.
A.O., V.M. and S.K. acknowledge funding by the Deutsche Forschungsgemeinschaft (DFG, German Research Foundation)-217133147/SFB 1073 (Project No. B07). S.K. also acknowledges partial support through National Science Foundation grants PHY-2210452, PHY-1748958 and PHY-2309135.

E.P. and A.O. contributed equally to this work.

\appendix 

\section{Effective Hamiltonian for the atomic limit}
\label{app:H_eff}

In this appendix we review the strong coupling expansion of the Holstein model (see also \cite{BeniPincusKanamori1974}\cite{Freericks1993}\cite{HirschFradkin1983}) and compute the resulting time scale estimate for the CDW order parameter factorization in this parameter regime.
In order to evaluate eq.~\eqref{eq:H_eff_gen}, we need to choose basis sets for the ground state manifold of $\hat H_{\rm ph} + \hat H_{\rm int}$ and its orthogonal complement, respectively. Under the condition of a half-filled lattice, the ground state manifold is bijective to the set of lattice bipartitions, i.e.~every bipartition $\mathcal L$ defines a valid ground state and vice versa,
\begin{equation}
    \ket{\Psi_0^{\mathcal L}} = \prod_{j\in \mathcal L} \hat C^\dagger_j \ket{0}_{\rm el} \ket{0}_{\rm ph}.
    \label{eq:Psi_0}
\end{equation}
The remaining states then belong to the orthogonal complement of the ground state manifold and are of the form
\begin{equation}
    \ket{\Psi_M^{\mathcal L}} = \prod_{j\in \mathcal L} \hat C^\dagger_j \ket{0}_{\rm el} \ket{m_j}_{\rm ph},
\end{equation}
where $M= \{m_j\}_j$ denotes the local phonon excitations for each site $j$, with $\sum_j m_j \neq 0$. First, we evaluate the term linear in $t_h$, which can only couple two ground states which differ by the position of exactly one electron. Let $\mathcal L$ and $\mathcal L'$ be such bipartitions, then
\begin{equation}
    \braket{\Psi_0^{\mathcal L}|\hat H_{\rm el}|\Psi_0^{\mathcal L'}} = -t_h |\braket{0|\mathrm e^{\frac{\gamma}{\omega_0}(\hat b^\dagger-\hat b)}|0}_{\rm ph}|^2 = -t_h \mathrm e^{-\gamma^2/\omega_0^2}.
\end{equation}
Hence, due to the nature of the ground states \eqref{eq:Psi_0}, the effective Hamiltonian up to linear order in $t_h$ takes the form
\begin{equation}\begin{split}
    \mathrm P \hat H \mathrm P &= \mathrm P( \hat H_{\rm ph} +\hat H_{\rm int} )\mathrm P + \mathrm P \hat H_{\rm el} \mathrm P \\
    &= -\frac{\gamma^2}{\omega_0} N_{\rm el} - t_h \mathrm e^{-\gamma^2/\omega_0^2} \sum_{\langle ij\rangle} \hat C^\dagger_i \hat C_j 
\end{split}
    \label{eq:H_eff_lin}
\end{equation}
As in the main text, $\langle ij\rangle $ (and, similarly, $\langle ijl\rangle $) denotes an ordered pair (or triple), e.g.~$\langle 1\,2\rangle \neq \langle 2\,1\rangle $.

For the part quadratic in $t_h$, we can distinguish two processes: Either one electron moves two sites away from its starting site, assuming the hopping processes are not forbidden by the Pauli exclusion principle, or the electron moves to the neighboring site and in the second hopping process moves back to the original site.
The first process we term “correlated second-nearest neighbor hopping” (corr.hop), whereas the second process gives rise to a repulsive nearest neighbor density-density interaction (nn).
In case of the correlated hopping, the matrix elements 
take the form
\begin{equation}\begin{split}
    &(\hat H_{\rm eff,corr.hop})_{\mathcal L,\mathcal L'} =\\
    &\quad +t_h^2 \sum_{\stackrel{l\in\rm NN(j)}{\cup \rm NN(j')}} \braket{\Psi_0^{\mathcal L} |  \hat c^\dagger_{j} \hat c_{l} \left( \sum_{\mathcal K, M} \frac{\ket{\Psi_M^{\mathcal K}}\bra{\Psi_M^{\mathcal K}}}{-\omega_0 \sum_i m_i} \right) \hat c^\dagger_{l} \hat c_{j'} | \Psi_0^{ \mathcal L'}},
\end{split} \label{eq:H_cnnn}
\end{equation}
where $M= \{m_i\}_i$ with $\sum_i m_i \neq 0$, as argued before.
W.l.o.g.~the (multi-)index of the initial and final lattice site of this correlated hopping process is taken to be $j'$ and $j$, respectively. That is, we assumed that these are the sites in which the bipartitions $\mathcal L$ and $\mathcal L'$ differ from each other. As becomes evident from eq.~\eqref{eq:H_cnnn}, the lattice sites $j$ and $j'$ have to be next-nearest neighbors. We evaluate this expression for one specific hopping path, i.e.~for fixed intermediate site $l$. Employing the relations $\braket{0|\mathrm e^{-\frac{\gamma}{\omega_0}(\hat b^\dagger - \hat b)}|n}_{\rm ph} = \left( \frac{\gamma}{\omega_0} \right)^n \mathrm e^{-\gamma^2/2\omega_0^2} / \sqrt{n!}$ and $\braket{0|\mathrm e^{\frac{\gamma}{\omega_0}(\hat b^\dagger - \hat b)}|n}_{\rm ph} = (-1)^n \braket{0|\mathrm e^{-\frac{\gamma}{\omega_0}(\hat b^\dagger - \hat b)}|n}_{\rm ph}$, we find
\begin{equation}\begin{split}
    &(\hat H_{\rm eff,corr.hop})_{\mathcal L,\mathcal L'} \Big|_{l\,\rm fixed} \\
    \quad &= -\frac{t_h^2}{\omega_0} \sum_{m=1}^\infty \frac{|\braket{0|\mathrm e^{\frac{\gamma}{\omega_0}(\hat b^\dagger - \hat b)}|m}_{\rm ph}|^2 \, |\braket{0|\mathrm e^{\frac{\gamma}{\omega_0}(\hat b^\dagger - \hat b)}|0}_{\rm ph}|^2}{m}\\
    &= -\frac{t_h^2}{\omega_0} \sum_{m=1}^\infty \left(\frac{\gamma}{\omega_0}\right)^{2m} \frac{\mathrm e^{-2\gamma^2/\omega_0^2} }{m! m} \\
    &= -\frac{t_h^2}{\omega_0} \mathrm e^{-2\gamma^2/\omega_0^2} \, I(\gamma^2/\omega_0^2) =: - t_2.
\end{split}\end{equation}
In the last step we inserted the identity $I(y):= \sum_{m=1}^\infty \frac{y^m}{m!\,m} = \sum_{m=1}^\infty \frac{y^m}{m!} \int_0^\infty \mathrm e^{-mk}\,\mathrm d k  = \int_0^\infty \sum_{m=1}^\infty \frac{\left(y\,\mathrm e^{-k}\right)^m}{m!} \,\mathrm d k = \int_0^\infty \left( \mathrm e^{y\,\mathrm e^{-k}} -1 \right) \, \mathrm d k = \int_0^{y} \frac{\mathrm e^x -1}{x} \,\mathrm d x$. With that, the correlated next-nearest neighbor hopping term in the effective Hamiltonian reads
\begin{equation}\begin{split}
    \hat H_{\rm eff,corr.hop} & = - t_2 \sum_{\stackrel{\langle ijl\rangle}{i\neq l}} \hat C_{l}^\dagger \hat C_{j} \hat C^\dagger_{j} \hat C_i \\
    &= - t_2 \sum_{\stackrel{\langle ijl\rangle}{i\neq l}} \hat C_{l}^\dagger (1-\hat n^{\rm el}_{j}) \hat C_i ,
\end{split} \label{eq:H_eff,corrhop}
\end{equation}
where one can identify the polaron number as the electron number, $\hat C_j^\dagger \hat C_j = \hat n_j^{\rm el}$.
Similiarly, we proceed with the remaining term of second order in $t_h$. This time, only diagonal matrix elements (i.e.~$\mathcal L = \mathcal L'$) contribute. Again, fixing the intermediate site $l\in\rm NN(j)$ in the two-step hopping process, we find
\begin{equation}\begin{split}
    (& \hat H_{\rm eff,nn})_{\mathcal L, \mathcal L} \Big|_{l\,\rm fixed}\\
   \quad &= t_h^2 \, \bra{\Psi_0^{\mathcal L}} \hat c_{j}^\dagger \hat c_{l} \left( \sum_{\mathcal K, M}  \frac{\ket{ \Psi_M^{\mathcal K} } \bra{ \Psi_M^{\mathcal K} } }{-\omega_0 \sum_i m_i} \right) \hat c_{l}^\dagger \hat c_{j} \ket{\Psi_0^{\mathcal L} }\\
   \quad &=-\frac{t_h^2}{\omega_0} \,\mathrm e^{-\frac{2\gamma^2}{\omega_0^2}}  I(2\gamma^2/\omega_0^2)=: - V_2/2.
\end{split} \label{eq:H_eff,nn}
\end{equation}
Again, $M= \{m_i\}_i$ with $\sum_i m_i \neq 0$ labels the site-specific phonon mode excitations. Without loss of generality the hopping starts from lattice site $j$. Ultimately this gives
\begin{equation}\begin{split}
    \hat H_{\rm eff,nn} &= -\frac{V_2}{2} \sum_{\langle l j \rangle}\, \hat C_{j}^\dagger \hat C_{l} \hat C^\dagger_{l} \hat C_j \\
    &= \frac{V_2}{2}  \sum_{\langle l j \rangle} \, \hat n^{\rm el}_{l} \hat n^{\rm el}_{j} - \frac{V_2 K_1}{2}N_{\rm el},
\end{split}
\end{equation}
where $K_1$ is the number of nearest neighbors in the lattice.
Neglecting any constant terms, the sum of eq.s \eqref{eq:H_eff_lin}, \eqref{eq:H_eff,corrhop} and \eqref{eq:H_eff,nn} gives the effective Hamiltonian \eqref{eq:H_eff} of the main text.
\begin{table}[t!]
            \renewcommand{\arraystretch}{1.6}
            \centering
            \begin{tabular}{l | c} \hline\hline \centering
                 \hspace*{2em} $t_1$ \hspace*{2em} &  $t_h \, \mathrm e^{-\gamma^2/\omega_0^2}$ \\ 
                 \hspace*{2em} $V_2$ \hspace*{2em} &  \hspace*{2em}  $2 \, t_h^2 \,\mathrm e^{-2\gamma^2/\omega_0^2} \, I(2\gamma^2/\omega_0^2) / \omega_0 $ \hspace*{2em} \\ 
                 \hspace*{2em} $t_2$ \hspace*{2em}  & $t_h^2  \mathrm e^{-2\gamma^2/\omega_0^2} \, I(\gamma^2/\omega_0^2) / \omega_0 $     \\ \hline\hline
            \end{tabular}
            \captionof{table}{Parameters of the effective Hamiltonian \eqref{eq:H_eff} for $\omega_0,\gamma\gg t_h$. $I(x) = \sum_{m=1}^\infty \frac{x^m}{m!\, m} = \int_{0}^x (\mathrm e^{y}-1)/y \, \mathrm d y $ and $I(x) \approx \mathrm e^x/x$ for $x\gg 1$, see also Tab.~\ref{tab:eff_params}.}
            \label{tab:eff_params_gen}
            \renewcommand{\arraystretch}{1}
\end{table}

In order to estimate the upper time limit for the validity of the factorization relation \eqref{eq:O_cdw_factorization} in the strong coupling regime- which is equivalent to the upper time limit until which the free polaron approximation remains justified- we perform a short time expansion of $\mathcal O_{\rm CDW}$ w.r.t.~$\hat H_{\rm eff}$. W.l.o.g.~we take the lattice site with \mbox{(multi)}index $0$ to host an electron at the initial time. Due to the translational symmetry of the charge occupation by two lattice sites in a periodic hypercubic lattice, we can rewrite $\mathcal O_{\rm CDW}(\delta t) = 2 \langle \hat n^{\rm el}_{0} (\delta t) \rangle  -1 $. Since $[\hat H_{\rm eff,nn},\hat n_0^{\rm el}]=0$ and $[\hat H_{\rm eff,corr.hop},\hat n_0^{\rm el}] \ket{\Psi_0^{\mathcal L_{\rm CDW}}} = 0$, the term linear in $\delta t$ follows a free-polaron time evolution, that is, w.r.t.~$\hat H_{\rm eff,hop}$ only. Similarly, one can show that the term $\sim \delta t^2$ also vanishes, which is simply due to the fact that both $\hat H_{\rm eff,nn}$ and $\hat H_{\rm eff,corr.hop}$ annihilate the initial state $\ket{\Psi_0^{\mathcal L_{\rm CDW}}}$, which, on top, is an eigenstate of $\hat n_0^{\rm el}$.
From the third order contribution, $ ((-i)^3 \delta t^3 /3)\, \left\langle  [\hat H_{\rm eff},[\hat H_{\rm eff},[\hat H_{\rm eff}, \hat n_0^{\rm el}]]] \right\rangle_{\Psi_0^{\mathcal L_{\rm CDW}}} $, the terms that deviate from noninteracting dynamics are
\begin{equation}\begin{split}
   & (-i)^3 \delta t^3 \Bigg( \left\langle \hat H_{\rm eff,hop} [\hat n_0^{\rm el}, \hat H_{\rm eff,corr.hop}]  \hat H_{\rm eff,hop} \right\rangle_{\Psi_0^{\mathcal L_{\rm CDW}}} \\
   & \quad + \frac{1}{3} \left\langle \hat H_{\rm eff,hop} [\hat n_0^{\rm el}, \hat H_{\rm eff,nn}]  \hat H_{\rm eff,hop} \right\rangle_{\Psi_0^{\mathcal L_{\rm CDW}}}  \Bigg).
\end{split} \label{eq:third-order-correction_factorization}
\end{equation}
We expect to observe free polaron dynamics in the order parameter as long as the absolute value of this term is much smaller than 1.

\section{Proof of the factorization relation of the order parameter}
\label{app:proof_factorization}

We consider a periodic, hypercubic lattice of dimension $d$; the lattice contains $L^d$ sites. We study the exact time evolution of the CDW order parameter under the Holstein Hamiltonian~\eqref{eq:Holstein_hamiltonian} with $\alpha=0$ in the no-coupling limit ($\gamma =0$). Each lattice site can be identified by the multi-index $l\equiv l_1\,l_2\,\dots\, l_d$, where each $l_i$ corresponds to the position along the $i$th dimensional axis. \\
Recall that the CDW order parameter is given by eq.~\eqref{eq:O_CDW_def}, which for $d$ dimensions ($d$D) results in
\begin{equation}
 \mathcal O_{\rm CDW}(t) = \langle \hat n^{\rm el}_{00\dots 0} (t) \rangle - \langle \hat n^{\rm el}_{10\dots 0} (t) \rangle,
\end{equation}
exploiting translational symmetry of the Hamiltonian and the initial CDW state by two lattice sites. W.l.o.g.~we have taken the lattice site with multi-index $l=00\dots 0$ to host an electron initially. Due to the checkerboard structure of the CDW state, we therefore have $ \mathcal L_{\rm CDW}^d := \{ l_1\dots l_d \,|\, (l_1 + l_2 + \dots + l_d ) \text{ is even} .\}$.
With the help of the dispersion relation of tight-binding electrons in one dimension, $ \epsilon_k := 2\,t_h \cos(2\pi k /L), $ we can write
\begin{equation}\begin{split}
 \langle &\hat n_{l_1\,l_2\,\dots\,l_d}(t)\rangle \\
     & = \frac{1}{L^2}  \sum_{\stackrel{m\in}{\mathcal L_{\rm CDW}^d}} \prod_{j=1}^d \sum_{\stackrel{k_j,k_j'}{ = 1}}^{L}  \mathrm e^{i\frac{2\pi}{L} (k_j-k'_j)(l_j-m_j)}\, \mathrm e^{it\,(\epsilon_{k_j}-\epsilon_{k'_j})}  .
\end{split}\end{equation}
Hence, in one dimension the CDW order parameter for $\gamma=0$ reads
\begin{align}\begin{split}
 \mathcal O_{\rm CDW}^{\gamma=0,1\rm D}  = & \sum_{\stackrel{m \in \mathcal L_{\rm CDW}}{\Leftrightarrow m \text{ even}}} \sum_{\stackrel{k,k'}{ = 1}}^L  \frac{1}{L^2} \mathrm e^{-i\frac{2\pi}{L}(k-k')m} \, \mathrm e^{it\,(\epsilon_k-\epsilon_{k'})} \\
 & \qquad  \times \left[1-\mathrm e^{i\frac{2\pi}{L}\,(k-k')}\right],
 \end{split}
 \label{eq:O_CDW_1D}
\end{align}
and for $d$ dimensions in general,
\begin{align}\begin{split}
    \mathcal O_{\rm CDW}^{\gamma=0,d\rm D} =& \sum_{\stackrel{m\in}{\mathcal L_{\rm CDW}^d}} \sum_{k,k'} \Bigg[ \prod_{j=1}^d  \frac{1}{L^{2}} \mathrm e^{-i\frac{2\pi}{L} (k_j-k'_j)m_j}  \mathrm e^{ it \,(\epsilon_{k_j}-\epsilon_{k'_j} )} \Bigg] \\
    & \qquad \times \left[1-\mathrm e^{i\frac{2\pi}{L}(k_1-k'_1)}\right].
\end{split}
\label{eq:O_CDW_nD}
\end{align}
The second sum runs over all lattice sites for each of the multi-indices $k$ and $k'$.

After these preliminaries, we state again the relation we want to prove in this appendix, the factorization of the order parameter in any dimension $d$,
\begin{equation}
\left(\mathcal O_{\rm CDW}^{\gamma=0,1\rm D}\right)^d = \mathcal O_{\rm CDW}^{\gamma=0,d\rm D}.
\end{equation}
The proof employs induction over the dimension $d$. The induction base ($d=1$) is trivial. Assume the relation above holds for some $(d-1)\in\mathbb N_{>0}$. Then consider the \mbox{$d$th} power of $\mathcal O_{\rm CDW}^{\gamma=0,1\rm D}$.\\
To simplify notation, the abbreviation $E_j^{m_j} := \exp\left(-i\frac{2\pi}{L}(k_j-k_j')m_j\right)$ will be used. After inserting the induction hypothesis we have
\begin{widetext}
\begin{align*}
    \left(\mathcal O_{\rm CDW}^{\gamma=0,1\rm D}\right)^{d} & = \mathcal O_{\rm CDW}^{\gamma=0,(d-1)\rm D} \cdot \mathcal O_{\rm CDW}^{\gamma=0,1\rm D} \\
    &= \sum_{k,k'} \Bigg[ \prod_{j=1}^{d} \frac{1}{L^{2}} \, \mathrm e^{ it \,(\epsilon_{k_j}-\epsilon_{k'_j} )}  \Bigg]  \left[1-\mathrm e^{i\frac{2\pi}{L}(k_1-k'_1)}\right] \left[1-\mathrm e^{i\frac{2\pi}{L}(k_{d}-k'_{d})}\right] \bigg( \sum_{\stackrel{m_{d}}{ \text{even}}} \sum_{\stackrel{m_1 m_2\dots m_{d-1}}{\in\mathcal L_{\rm CDW}^{d-1}}} E_1^{m_1} E_2^{m_2}\dots \, E_{d}^{m_{d}}  \bigg) \\
    &= \sum_{k,k'}  \Bigg[ \prod_{j=1}^{d} \frac{1}{L^{2}} \, \mathrm e^{ it \,(\epsilon_{k_j}-\epsilon_{k'_j} )}  \Bigg] \left[1-\mathrm e^{i\frac{2\pi}{L}(k_1-k'_1)}\right] \\
        & \qquad \qquad \times \Bigg[ \left(1-E_{d}^{-1}\right) \bigg( \sum_{\stackrel{m_1 m_2 \dots m_{d}}{\in\mathcal L^{d}_{\rm CDW}}} E_1^{m_1} \dots E_{d}^{m_{d}} - \sum_{\stackrel{m_{d}}{\text{odd}}} \sum_{\stackrel{m_1 m_2 \dots m_{d-1}}{\notin\mathcal L^{d-1}_{\rm CDW}}} E_1^{m_1} \dots E_{d}^{m_{d}}   \bigg)  \Bigg] \\
   &= \sum_{k,k'}  \Bigg[ \prod_{j=1}^{d} \frac{1}{L^{2}} \, \mathrm e^{ it \,(\epsilon_{k_j}-\epsilon_{k'_j} )}  \Bigg] \left[1-\mathrm e^{i\frac{2\pi}{L}(k_1-k'_1)}\right]  \\
   & \qquad \qquad \times  \Bigg[ \bigg( \sum_{\stackrel{m_1 m_2 \dots m_{d}}{\in\mathcal L^{d}_{\rm CDW}}} E_1^{m_1} \dots E_{d}^{m_{d}} \bigg) - \bigg( \sum_{m_{d} \text{ odd}} E_{d}^{m_{d}} \bigg) \underbrace{\bigg( \sum_{m_1 m_2 \dots m_{d-1}} E_1^{m_1} \dots E_{d-1}^{m_{d-1}}\bigg)}_{= \,\left( \sum_{m_1=1}^L E_1^{m_1}  \right)^{d-1} \quad \text{(I)}} \Bigg].
    \numberthis \label{eq:aux_derivation}
\end{align*}
\end{widetext}
For the last two equal signs one inserts the identity
\begin{equation}
    \sum_{\stackrel{m_1,\dots,m_{d}}{\in\mathcal L^{d}_{\rm CDW}}} \dots \, = \sum_{\stackrel{m_{d}}{\text{even}}} \sum_{\stackrel{m_1 m_2 \dots m_{d-1}}{\in\mathcal L^{d-1}_{\rm CDW}}} \hspace{-1em} \dots \quad + \sum_{\stackrel{m_{d}}{\text{odd}}} \sum_{\stackrel{m_1 m_2 \dots m_{d-1}}{\notin\mathcal L^{d-1}_{\rm CDW}}} \hspace{-1em}\dots \quad,
\end{equation}
which follows from the definition of $\mathcal L_{\rm CDW}^{d}$. We see that eq.~\eqref{eq:aux_derivation} is equal to $\mathcal O_{\rm CDW}^{\gamma=0,d\rm D}$ \eqref{eq:O_CDW_nD} if the term (I) vanishes. It contains the sum $\sum_{m_1 m_2 \dots m_{d-1}} \dots$, which is running over all lattice sites in a $(d-1)$-dimensional hyper-cubic lattice. We evaluate
\begin{equation}\begin{split}
    \sum_{m_1=1}^L E_1^{m_1} = \frac{1-\mathrm e^{-i\,2\pi(k_1 - k'_1)}}{1-\mathrm e^{-i\frac{2\pi}{L}(k_1 - k'_1)}} =
\begin{cases}
    \frac{L}{2} & \qquad \text{ if } k_1 = k'_1\,, \\
        0  & \qquad \text{ else,}    
\end{cases}                      
\end{split}\end{equation}
using l'Hospital's rule.
Now, we distinguish the cases $k_1\neq k_1'$ and $k_1= k_1'$. For $k_1\neq k_1'$, the term (I) vanishes. On the other hand, for $k_1\neq k_1'$, the second square bracket in eq.~\eqref{eq:aux_derivation}, $\left[1-\mathrm e^{i\frac{2\pi}{L}(k_1-k'_1)} \right] = 0$, hence this combination does not contribute to $\left(\mathcal O_{\rm CDW}^{\gamma=0,1\rm D}\right)^{d}$ anyway. Therefore, we are left with
\begin{equation}\begin{split}
    \left(\mathcal O_{\rm CDW}^{\gamma=0,1\rm D}\right)^{d} =& \sum_{k,k'}  \Bigg[ \prod_{j=1}^{d} \frac{1}{L^{2}} \, \mathrm e^{ it \,(\epsilon_{k_j}-\epsilon_{k'_j} )}  \Bigg] \left[1-\mathrm e^{i\frac{2\pi}{L}(k_1-k'_1)}\right] \\
    &\qquad \times \bigg( \sum_{\stackrel{m_1 m_2 \dots m_{d}}{\in\mathcal L^{d}_{\rm CDW}}} E_1^{m_1} \dots E_{d}^{m_{d}} \bigg) \\
    &= \mathcal O_{\rm CDW}^{\gamma=0,d\rm D}.  \hspace{12em}\hfill \square
\end{split}\end{equation}
\linebreak

In the remaining part of this appendix, we discuss the generalization of this factorization relation to other bipartite Bravais lattices, in particular the primitive orthorhombic and mono-/triclinic ones. Of course, other bipartite lattices, e.g.~the honeycomb lattice, may also exhibit a CDW. However, for these geometries, the calculation of the order parameter dynamics after the quench requires a slightly different approach and cannot be inferred directly from the computation above.

We focus on the tight-binding model in mono-/triclinic and orthorhombic lattices, where next-nearest neighbor hopping and other interaction effect are neglected. Then the Holstein Hamiltonian \eqref{eq:Holstein_hamiltonian} accounts for the lattice geometry only by a change in the hopping integral $t_h$, which will become anisotropic. The dispersion relation has to be altered accordingly, e.g.~$\epsilon(k_i,k_j,k_l) = 2 t^i_h \cos(2\pi k_i/L) + 2 t^j_h \cos(2\pi k_j/L)+2 t^l_h \cos(2\pi k_l/L) $ in 3D, where the super/subscripts $i,j$ and $l$ label the (reciprocal) primitive lattice vectors.

Due to the translational symmetries in primitive triclinic and orthorhombic lattices, the lattice direction along which the charge density difference is measured does not matter; the order parameter (unlike the dispersion) is isotropic. This is due to the fact that one can relate the unoccupied and occupied sites of the CDW via symmetry transformations. Therefore we have $\mathcal O^{\gamma=0,3\mathrm{D}}_{\mathrm{CDW},\bm i}(t) = \mathcal O^{\gamma=0,3\mathrm{D}}_{\mathrm{CDW},\bm j}(t) = \mathcal O^{\gamma=0,3\mathrm{D}}_{\mathrm{CDW},\bm l}(t)=:\mathcal O^{\gamma=0,3\mathrm{D}}_{\mathrm{CDW}}(t)$, where, again, $i,j$ and $l$ correspond to the primitive lattice vectors. The same holds for two dimensions and so on.
Making use of the new dispersion relation, the proof of the factorization relation can be carried out in the same manner as above. Ultimately, one finds
\begin{equation}
    \mathcal O^{\gamma=0,3\mathrm{D}}_{\mathrm{CDW}}(t) = \mathcal O^{\gamma=0,1\mathrm{D}}_{\mathrm{CDW},\bm i}(t) \cdot \mathcal O^{\gamma=0,1\mathrm{D}}_{\mathrm{CDW},\bm j}(t) \cdot \mathcal O^{\gamma=0,1\mathrm{D}}_{\mathrm{CDW},\bm l}(t),
\end{equation}
which is a generalization of eq.~\eqref{eq:O_cdw_factorization} for primitive orthorhombic and triclinic lattices.\\
This factorization breaks down as soon as electronic tunneling between next-nearest neighbor sites has to be included. In general, the order parameter can also become anisotropic and one needs to distinguish between the different lattice directions.

\section{Calculation of $x_i^\text{ph}(t)$ from $\mathcal{O}_\text{CDW}(t)$\label{app:x_ph}}

The operators
\begin{align}\begin{split}
\hat x_i^\text{ph} = \frac{1}{\sqrt{2 \omega_0}} \big( \hat b_i + \hat b_i^\dagger \big) ~~\text{and}~~ \hat p_i^\text{ph} = i \sqrt{\frac{\omega_0}{2}} \big( \hat b_i^\dagger - \hat b_i \big)
\end{split}\end{align}
obey the following equations of motion:
\begin{align}\begin{split}
\partial_t \hat x_i^\text{ph} &= \hat p_i^\text{ph}, \quad \partial_t \hat p_i^\text{ph} = \sqrt{2 \omega_0} \gamma \big( \hat n_i - \alpha \big) - \omega_0^2 \hat x_i^\text{ph} .
\end{split}\end{align}
Combining the two equations yields
\begin{align}\begin{split}
\partial_t^2 \hat x_i^\text{ph} &= \sqrt{2\omega_0} \gamma \big( \hat n_i - \alpha \big) - \omega_0^2 \hat x_i^\text{ph} .
\end{split}\end{align}

In the following, we set $\alpha = 0$ and focus on the dynamics after the hopping quench, leading to the CDW melting.
Let us consider a CDW unit cell with $\hat x_0^\text{ph}$ on the initially occupied and $\hat x_1^\text{ph}$ on the initially unoccupied site.
Define $\hat x_+^\text{ph} := \hat x_0^\text{ph} + \hat x_1^\text{ph}$, $\hat x_-^\text{ph} := \hat x_0^\text{ph} - \hat x_1^\text{ph}$, $x_\pm^\text{ph} = \langle \hat x_\pm^\text{ph} \rangle$, $n_0 + n_1 = 1$, such that
\begin{align}\begin{split}
\partial_t^2 \hat x_+^\text{ph} &= \sqrt{2 \omega_0} \gamma \big( \hat n_0 + \hat n_1 \big) - \omega_0^2 \hat x_+^\text{ph} , \\
\rightarrow\quad \partial_t^2 x_+^\text{ph} &= \sqrt{2 \omega_0} \gamma  - \omega_0^2 x_+^\text{ph} , \\
\partial_t^2 \hat x_-^\text{ph} &= \sqrt{2 \omega_0} \gamma \big( \hat n_0 - \hat n_1 \big) - \omega_0^2 \hat x_-^\text{ph} ,\\
\rightarrow\quad \partial_t^2 x_-^\text{ph} &= \sqrt{2 \omega_0} \gamma \mathcal{O}_\text{CDW}(t) - \omega_0^2 x_-^\text{ph} .
\end{split}\end{align}

With the initial conditions $x_0^\text{ph}(0) = \sqrt{2} \gamma/\omega_0^{3/2}$, $x_1^\text{ph}(0) = 0$, hence $x_+^\text{ph}(0) = x_-^\text{ph}(0) =x_0^\text{ph}(0)= \sqrt{2} \gamma/\omega_0^{3/2}$ and $\partial_t x_\pm^\text{ph}(0) = 0$, the solutions to the differential equations read
\begin{align}\begin{split}
 &x_+^\text{ph}(t) = \sqrt{2} \frac{\gamma}{\omega_0^{3/2}} , \\
 &x_-^\text{ph}(t) = \cos(\omega_0 t) \sqrt{2} \gamma/\omega_0^{3/2} \\
 &\quad + \sqrt{2} \gamma/\omega_0^{1/2} \sin(\omega_0 t) \int_0^t \mathcal{O}_\text{CDW}(\tau) \cos(\omega_0 \tau) \text{d}\tau \\
 &\quad - \sqrt{2} \gamma/\omega_0^{1/2} \cos(\omega_0 t) \int_0^t \mathcal{O}_\text{CDW}(\tau) \sin(\omega_0 \tau) \text{d}\tau ,
\end{split}\end{align}
which gives equations \eqref{eq:x_dyn} stated in the main text.

If the electronic order parameter in one spatial dimension stays close to one, $\mathcal{O}_\text{CDW}(t) = 1 - \epsilon(t)$, one can make use of $\big( \mathcal{O}_\text{CDW}(t) \big)^d = \big( 1 - \epsilon(t) \big)^d = 1 - d \cdot \epsilon(t) + \mathcal{O}(\epsilon^2)$ which yields
\begin{align}\begin{split}
 x_1^\text{ph}(t) &= \frac{\gamma d}{\sqrt{2}\omega_0^{3/2}} \Big( \sin(\omega_0 t) \int_0^t \epsilon(\tau) \cos(\omega_0 \tau) \text{d}\tau \\
 &\qquad - \cos(\omega_0 t) \int_0^t \epsilon(\tau) \sin(\omega_0 \tau) \text{d}\tau \Big)
\end{split}\end{align}
\newline

\bibliography{lit.bib}

\end{document}